\documentstyle[amsfonts,multicol,fancyheadings,eqsecnum,prd,aps]%
{revtex}

\def\buildchar#1#2#3{\null \! \mathop {\vphantom {#1}\smash
#1}\limits ^{#2}_{#3}\!\null }
\def\OT#1{\buildchar{{#1}}{\;_\sim}{}\/}
\def\UT#1{\buildchar{{#1}}{}{^\sim}\/}
\def\OTT#1{\buildchar{{#1}}{\;_\approx}{}\/}
\def\UTT#1{\buildchar{{#1}}{}{^\approx}\/}

\begin{document}
\draft


\title{Gauge group and reality conditions in Ashtekar's complex
formulation of canonical gravity}

\author{J.\ M.\ Pons
\footnote[1]{Electronic address: pons@ecm.ub.es}}
\address{Departament d'Estructura i Constituents de la Mat\`eria, 
Universitat de Barcelona,\\
and Institut de F\'\i sica d'Altes Energies,\\
Diagonal 647, E-08028 Barcelona, Catalonia, Spain}

\author{D.\ C.\ Salisbury 
\footnote[2]{Electronic address: dsalisbury@austinc.edu}} 
\address{Department of Physics, 
Austin College, Sherman, Texas 75090-4440, USA}

\author{L.\ C.\ Shepley 
\footnote[3]{Electronic address: larry@helmholtz.ph.utexas.edu}} 
\address{Center for Relativity, Physics Department, \\
The University of Texas, Austin, Texas 78712-1081, USA \\ ~}

\date{Version of 22 December 1999\ --- to be submitted to 
\textit{Phys.\ Rev.}}
\maketitle


\begin{abstract}

We discuss reality conditions and the relation between spacetime 
diffeomorphisms and gauge transformations in Ashtekar's complex 
formulation of general relativity.  We produce a general theoretical 
framework for the stabilization algorithm for the reality conditions, 
which is different from Dirac's method of stabilization of 
constraints.  We solve the problem of the projectability of the 
diffeomorphism transformations from configuration-velocity space to 
phase space, linking them to the reality conditions.  We construct the 
complete set of canonical generators of the gauge group in the phase 
space which includes all the gauge variables.  This result proves that 
the canonical formalism has all the gauge structure of the Lagrangian 
theory, including the time diffeomorphisms.

\end{abstract}

\pacs{04.20.Fy, 11.10.Ef, 11.15.-q 
\hfill gr-qc/9912085}

\setlength{\columnseprule}{0pt}\begin{multicols}{2}


\section{Introduction}
\label{sec:intro}
  
In recent papers \cite{pons/salisbury/shepley/97,%
pons/salisbury/shepley/99a,pons/salisbury/shepley/99b} we have 
discussed some special features exhibited by the gauge groups in 
Einstein and Einstein-Yang-Mills theories and in a real triad approach 
to general relativity when their formulations are brought from 
configuration-velocity space (the tangent bundle $TQ$) to phase space 
(the cotangent bundle $T^{*}\!Q$).  Our viewpoint is that the 
configuration-velocity space and phase space formulations are 
equivalent (see \cite{pons/shepley/95}).  We found that some of the 
generators of the diffeomorphism group in the tangent bundle are not 
projectable to the cotangent bundle.  To make them projectable, the 
otherwise arbitrary functions in the gauge group generators must 
depend on the field variables, particularly on the lapse function and 
shift vector of the metric---though this dependence still allows all 
infinitesimal diffeomorphisms to be represented.  In 
Einstein-Yang-Mills and triad theories, diffeomorphisms must be 
accompanied by other gauge transformations in order to be projectable.  
When projectability is achieved, we have the full proof that indeed 
the gauge group is the same in configuration-velocity space as in 
phase space; this identity of the gauge group is not widely 
recognized.

Here we study in detail the issue of the gauge group in the Ashtekar 
complex formulation \cite{ashtekar/86,ashtekar/87,ashtekar/91} of 
canonical gravity.  Ashtekar's use of a self-dual connection makes 
this formulation very similar to a Yang-Mills theory, and so we expect 
to get and do get results similar to our previous results.  However, a 
somewhat unusual aspect of this program is the use of a complex 
Lagrangian and a complex Hamiltonian.  The fact that Ashtekar's 
connection is complex introduces essential novelties.  To recover real 
gravity, reality conditions must be imposed, and we make a thorough 
examination of them.  These conditions are not constraints in a Dirac 
sense \cite{dirac/50,dirac/64}.  We develop the theoretical framework 
for a stabilization algorithm to maintain the reality conditions under 
time evolution.  This algorithm is different from the Dirac 
stabilization algorithm for constraints because of the complex 
character of the Hamiltonian, though our treatment is conceptually 
close to Dirac's method.

Recently generalization's of Ashtekar's complex formalism have been 
introduced.  In one approach it has been shown that general relativity 
can be reformulated as a one-parameter family of real connections 
\cite{barbero/95,holst/96,immirzi/97}.  When the otherwise real 
parameter takes the value $i$, one recovers the Ashtekar complex 
connection.  However, one apparent drawback to this real approach is 
that the scalar constraint loses the simple form it assumes in the 
complex regime.  This could constitute a serious obstacle for the 
quantization program, though it is true that difficulties in 
constructing a Hilbert space satisfying the reality conditions in the 
complex Ashtekar program are thereby circumvented.  A second approach 
undertakes a generalized Wick transform of the complex connection to a 
real connection \cite{thiemann/96,ashtekar/96}.  This transform has 
been shown under certain circumstances to be equivalent to an analytic 
continuation to imaginary time \cite{mena/97}, and thus to a 
spacetime with Riemannian signature.  The advantage one hopes to gain 
through this transform is that it may be possible to solve the simpler 
scalar constraint in the Lorentzian sector and then implement the Wick 
transform, thus satisfying the reality conditions.

The argument we put forth here is that the relevance of the complex 
Ashtekar approach has certainly not diminished.  A major theme in this 
paper is the relation of the scalar constraint to spacetime 
diffeomorphisms.

Our purposes in this paper are twofold: On the one hand, we will 
clarify the structure of the generators of the gauge group in the 
complex Ashtekar formulation of canonical gravity.  On the other hand, 
we will discuss fully the stabilization algorithm for the reality 
conditions.  It is not surprising---perhaps---that both aspects, gauge 
group and reality conditions, are related: Any symmetry, including 
gauge symmetries must preserve the reality conditions.  We will 
exhibit the links that exist between these conditions and the 
conditions of projectability from configuration-velocity to phase 
space of gauge variations.  We distinguish between metric reality 
conditions (only the full spacetime metric itself must be real) and 
triad reality conditions (the spatial orthonormal triad vectors, as 
well as the metric, must be real) as in 
\cite{bergmann/smith/91,yoneda/shinkai/96}.  We will see that the 
rotation gauge group (for the triads) is reduced from $SO(3,C)$ to 
$SO(3,R)$ to fulfill the triad reality conditions.  Our results 
concerning the reality conditions do agree with those of 
\cite{yoneda/shinkai/96}; our contribution is that we make clear when 
the stabilization algorithm for the reality conditions is terminated 
and how it applies in a general sense.  Also, we give a thorough 
discussion of the elimination of part of the gauge freedom when we 
extend reality conditions from metric to triad.

We explicitly assume that the connection $A^{i}_{\mu}$ is complex but 
also consider the possibility that all variables in phase space are 
complex.  It is significant that all the gauge variables, that is the 
lapse , the shift , and the time component of the connection 
$A^{i}_{0}$, are retained as canonical variables in the analysis of 
gauge symmetries which we will present.  In particular, it could well 
prove useful in quantum gravity to retain $A^{i}_{0}$ as an operator.  
We would thus contemplate holonomies, parallel transporters of SU(2), 
in directions off the constant-time hypersurfaces.  We presume that 
all functions, including the Hamiltonian, are analytic, and that phase 
space has a standard Poisson bracket structure.  Physical reasons 
require that some of the variables must be real.  Then it is necessary 
to impose restrictions on the initial conditions and to restrict gauge 
freedom in such a way that time evolution will keep real these 
variables.  These restrictions are called the reality conditions.

This paper is organized as follows: The stabilization algorithm for 
the reality conditions is presented in Section \ref{sec:genreal}.  The 
algorithm is general in the sense that it can be applied to any 
complex theory in which physical reasons require that some of the 
variables be real.  In Section \ref{sec:lag}, the Ashtekar approach is 
succinctly introduced with some results and notations.  The canonical 
approach is undertaken in Section \ref{sec:can}, and in Section 
\ref{sec:real} we apply the reality condition algorithm to the case of 
Ashtekar canonical gravity.  In Sections \ref{sec:proj} and 
\ref{sec:gen} we solve the problem of finding the projectable gauge 
transformations and their canonical generators, finding in the process 
some interesting relations with the reality conditions.  We discuss 
the counting of degrees of freedom in Section \ref{sec:count}.  We 
devote Section \ref{sec:conc} to conclusions.


\section{Stabilization algorithm for reality conditions---general 
theory}
\label{sec:genreal}

In this Section we provide the theoretical setting for what properly 
must be called the stabilization algorithm for the reality conditions.  
This setting is applicable to any dynamical theory that makes use of 
complex variables but requires that some of these variables be real to 
be physically acceptable.  In other words, initial conditions must fix 
real values for these variables, and time evolution must preserve the 
reality.

Reality conditions are not constraints in the Dirac sense.  The 
difference comes from the fact that reality conditions do not place 
restrictions on the variables of the formalism but only on the values 
of some real or imaginary parts of these variables.  The difference is 
made even more clear when we consider stabilization procedures.  If 
the Dirac Hamiltonian is, say, $H$, the stabilization of a (time 
independent) Dirac-type constraint $\phi$ is to require the tangency 
of the dynamical vector field $\{-,\,H\}$ on the surface defined by 
$\phi=0$:
\[  \{\phi,\,H\} = 0 \ .
\]
This requirement may introduce new constraints or the determination 
some arbitrary functions in $H$.  The stabilization of a Dirac 
constraint follows this procedure whether $H$ is real or complex.

Instead, if we have a (time independent) reality condition, such as 
the vanishing of the imaginary part of a quantity $f$, $\Im f=0$, its 
stabilization involves, at least, the requirement
\[  \Im\{f,\,H\} = 0 \ .
\]
This is not a tangency condition.  Moreover, the expression 
\[   \{\Im f,\,H\}
\]
makes no sense at all in the formalism, because the bracket is defined 
for complex phase space variables and cannot be applied to real or 
imaginary parts of these variables.

Before developing the correct stabilization for reality conditions, we 
briefly review the basics of the stabilization algorithm for Dirac 
constraints.  Similarities and differences between the two 
stabilization procedures will become evident.


\subsection{Stabilization of Dirac constraints}
\label{subsec:stab.Dirac}

Dirac's method applies both to the Lagrangian and Hamiltonian 
formalisms, but here we will only consider its implementation in the 
latter case.  Consider a dynamical evolution in phase space with some 
gauge freedom.  We start with the canonical Hamiltonian $H_{\rm c}$, 
whose pullback to configuration-velocity space is the Lagrangian 
energy
\begin{equation}
    E_{L} := \dot q^{i} {\partial L\over\partial\dot q^{i}} - L \ ,
        \label{eq:lageng}
\end{equation}
where $L$ is the Lagrangian, which we take to be time-independent, 
$\{q^{i}\}$ are the configuration components, and $\dot{}$ is $d/dt$.  
The Dirac Hamiltonian is
\[  H_{\rm D} = H_{\rm c} + \lambda^{\mu}\phi_{\mu}\ ; 
\]
the $\phi_{\mu}$ are the primary constraints, ${\mu}=1,\ldots,n$, and 
$\lambda^{\mu}$ are Lagrange multipliers (arbitrary functions in 
principle) that describe the gauge freedom available to this system.  
The first step in Dirac's method is to ask for the dynamics to result 
in trajectories tangent to the primary constraint surface.  This 
requirement of tangency may lead to the determination of some of the 
multipliers $\lambda^{\mu}$ and the appearance of new constraints.  
The next step is again to require that the trajectories be tangent to 
the new constraint surface.  The stabilization procedure continues and 
eventually is completed.

We analyze this procedure from the point of view of finite time 
evolution for application in Subsection \ref{subsec:stab.real}.  To 
make things simpler, as an example, we assume that none of the 
multipliers $\lambda^{\mu}$ are determined at any step of the above 
procedure.  Then, as far as the time-evolution of the constraints is 
concerned, we can use the time-independent $H_{\rm c}$ as the 
dynamical generator.  We start with the primary constraints 
$\phi_{\mu}$.  The time evolution operator from time zero to time $t$ 
is
\begin{equation}
    E[t] = \exp (t \{-, \, H_{\rm c} \})\ ,
        \label{evol-op}
\end{equation}
with the expansion 
\begin{eqnarray}
    \phi_{\mu}[t] 
    &=& E[t] \phi_{\mu}  \nonumber \\
    &=& \phi_{\mu} + t \{\phi_{\mu},\,H_{\rm c}\} 
        + {t^{2}\over 2}  \{\{\phi_{\mu},\,H_{\rm c}\},\,H_{\rm c}\} 
            \nonumber\\
    &&\quad + {t^{3}\over3!}
        \{\{\{\phi_{\mu},\ H_{\rm c}\},\,H_{\rm c}\},\,H_{\rm c}\}
        +\ldots
                \nonumber \\
    &=:& \sum_{n=0}^{\infty}{t^{n}\over n!}
        \{\phi_{\mu},\,H_{\rm c}\}_{(n)} \ ;
                \label{expansion}
\end{eqnarray}
in this expression $\phi_{\mu}[t]$ is the function $\phi_{\mu}(x(t))$, 
where $x(t):=(q(t),p(t))$ is the trajectory in phase space satisfying 
the equations
\[  {\dot x}(t) = \{x,\,H_{\rm c}\}|_{x=x(t)} \ .
\]
To preserve the primary constraints under finite evolution we must 
require
\[  \phi_{\mu}[t] = 0
\]
for any $t$. This is the same as the infinite set of restrictions 
\begin{equation}
    \{\phi_{\mu},\,H_{\rm c}\}_{(n)}=0 \ ;
        \label{conset}
\end{equation}
note that $n=0$ corresponds to the primary constraints $\phi_\mu=0$.

In general, the $n=1$ level of stabilization in (\ref{conset}), 
$\{\phi_{\mu},\,H_{\rm c}\}=0$, may introduce new independent 
constraints (secondary constraints) 
$\phi_{\mu}^{(1)}:=\{\phi_{\mu},\,H_{\rm c}\}$.  The second level of 
stabilization is $\{\phi_{\mu}^{(1)},\,H_{\rm c}\}=0$, which is 
Dirac's requirement that the vector field $\{-,\,H_{\rm c}\}$ be 
tangent to the new constraint surface (defined by all the primary and 
secondary constraints).  It is worth noticing that in general the 
algorithm to get new constraints will eventually stop, and only a 
finite number of the requirements in (\ref{conset}) will be relevant.

For instance, if there are no tertiary constraints, the $n=2$ level of 
stabilization is satisfied when the primary and secondary constraints 
are taken into account.  Then, $\{\phi_{\mu}^{(1)},\ H_{\rm c}\}$ is a 
linear combination of the primary and secondary constraints.  All 
other terms in (\ref{conset}) vanish under the condition that all of 
the primary and secondary constraints are satisfied.  There are 
exceptions to this casual statement, in particular when some of the 
constraints are not effective (an effective constraint has 
nonvanishing differential on the constraint surface), and we discuss 
them in the next Subsection.  With these exceptions, the stabilization 
procedure terminates when we find a level of stabilization that is 
already satisfied under the requirements introduced in the previous 
levels.

The general situation is when we must consider time dependence in 
$H_{D}$ (because of the $\lambda^{\mu}$).  In this case, 
$H_{\rm D}(t_{1})$ does not necessarily have vanishing Poisson 
bracket with $H_{\rm D}(t_{2})$, for $t_{1}\neq t_{2}$.  The time 
evolution operator (\ref{evol-op}) is then replaced by
\begin{equation}
    E[t] 
    ={\cal T} \exp(\int_{0}^{t}dt'\{-,\,H_{\rm D}(t')\}) \ ,
        \label{evol-op2}
\end{equation} 
where ${\cal T}$ is the time-ordering operator: It acts as 
\[  {\cal T}\{\{-,\, H_{\rm D}(t_{1})\},\, H_{\rm D}(t_{2})\} 
    = \{\{-,\, H_{\rm D}(t_{<})\}, \, H_{\rm D}(t_{>})\} \ ,
\]
with $t_{>}=\max(t_{1},t_{2})$ and $t_{<}=\min(t_{1},t_{2})$ 
(this expression generalizes to any order).
 
The levels of stabilization in (\ref{evol-op2}) now become
\begin{eqnarray}
    \{\phi_{\mu},\,H_{\rm D}(t)\} 
        &=& 0 \nonumber \\
    \{\{\phi_{\mu},\,H_{\rm D}(t_{1})\},\,H_{\rm D}(t_{2})\} 
        &=& 0 \nonumber \\
    \{\{\{\phi_{\mu},\, H_{\rm D}(t_{1})\},\,H_{\rm D}(t_{2})\},
        H_{\rm D}(t_{3})\}&=& 0 \nonumber\\
    \ldots &&  ,
            \label{recurr2}
\end{eqnarray}
with $t_{1}<t_{2}<t_{3}<\ldots$.  These requirements (\ref{recurr2}) 
may determine some of the arbitrary functions in $H_{\rm D}$ or they 
may bring forth further constraints.  Once an arbitrary function gets 
determined, it can be replaced by its expression in phase space for 
all remaining levels of stabilization.

The sequence (\ref{recurr2}) eventually terminates when the 
stabilization equations for all the constraints no longer determine 
new constraints: Higher stabilization equations are automatically 
satisfied.


\subsection{An aside on ineffective constraints}
\label{subsec:stab.ineff}

There is an exception to the rule, just enunciated, that says that the 
stabilization algorithm is finished when, at a given level, no new 
constraints appear.  The expression $\{\phi_{\mu}^{(1)},\,H\}=0$ is 
meant to be Dirac's requirement that the vector field $\{-,\,H\}$ be 
tangent to the constraint surface defined by the primary and secondary 
constraints.  This is not an accurate statement when a secondary 
constraint is ineffective (the primary constraints are always taken in 
effective form), that is, if its differential vanishes on the 
constraint surface.  For instance, consider the effective constraint 
$\phi$.  To make it ineffective we can square it to get $f=\phi^{2}$.  
The two constraints still define the same surface, 
$\phi=0\Longleftrightarrow f=0$.  However, the vanishing of 
$\{f,\,H\}$ does not imply the tangency of $\{-,\,H\}$ to the surface 
$f=0$ but rather a triviality, because $\{f,\,H\}=2\phi\,\{\phi,\,H\}$ 
automatically vanishes on $f=0$.  This reflects the ineffective 
character of $f$ (but notice that $\{f,\,H\}$ cannot be expressed as a 
linear combination of $f$ with the coefficient being regular at the 
surface $f=0$).

Because of the possible presence of ineffective constraints, it may be 
true that one level of stabilization does not bring new restrictions, 
and yet subsequent levels do.  In fact, in our example with $f$ 
ineffective, the next level of stabilization produces 
$\{\{f,\,H\},H\}=2\phi\{\{\phi,\,H\},H\}+2\{\phi,\,H\}^{2}$.  This 
could introduce a new ineffective constraint $\{\phi,\,H\}^{2}=0$ that 
defines the same surface as $\{\phi,\,H\}=0$.

The moral is that if we have ineffective constraints, we must take 
special precautions that the tangency conditions are correctly 
implemented and that all levels of equation (\ref{recurr2}) are 
examined.


\subsection{Stabilization of reality conditions}
\label{subsec:stab.real}

Suppose that our reality condition requires that the functions 
$f_{\alpha}$, for some set of indices ${\alpha}$, must be kept real 
under time evolution.  We begin, for simplicity, with the case when 
the Lagrangian multipliers play no part, as in Subsection 
\ref{subsec:stab.Dirac}; then we may work with the time-independent 
canonical Hamiltonian $H_{\rm c}$.  Expressed in the notation 
introduced above, the reality requirement is
\[  \Im(f_{\alpha}[t]) = 0 \ ,
\] 
which is, using the evolution operator (\ref{expansion}), 
\begin{eqnarray}
    \Im(f_{\alpha}[t]) 
    &=& \Im(E[t]f_{\alpha})  
            \nonumber\\
    &=& \sum_{n=0}^{\infty}{t^{n}\over n!}
        \Im\{f_{\alpha},\,H_{\rm c}\}_{(n)} =0 \ ,
                \label{expansion2}
\end{eqnarray} 
for any $t$.  Therefore, in addition to the primary reality condition, 
\[ \Im f_{\alpha} = 0\ ,
\] 
we get the levels of stabilization
\begin{eqnarray}
    \Im\{f_{\alpha},\,H_{\rm c}\} 
        &=& 0 
            \nonumber\\ 
    \Im\{\{f_{\alpha},\,H_{\rm c}\},\,H_{\rm c}\} 
        &=&0 
            \nonumber\\ 
    \Im\{\{\{f_{\alpha},\,H_{\rm c}\},\,H_{\rm c}\},\,H_{\rm c}\}
        &=&0 
            \nonumber\\ 
    \ldots 
        & & \ .
            \label{recurr3}
\end{eqnarray}
We call these conditions the secondary reality condition, tertiary 
reality condition, and so on.  Notice in fact that all these 
requirements need only to hold on the constraint surface, because the 
complete dynamical setting is given by the evolution operator 
(\ref{evol-op}) supplemented with the Dirac constraints.

One striking difference between these conditions (\ref{recurr3}) and 
the Dirac stability conditions (\ref{recurr2}) is that the vanishing 
of one level of stabilization due to the fulfillment of the previous 
ones does not guarantee that the subsequent levels will also vanish.  
For instance, let us suppose that
\[  \Im\{f_{\alpha},\,H_{\rm c}\} 
    = \eta^{\beta}_{\alpha}\, \Im(f_{\beta}) \ , 
\]
for a real matrix $\eta^{\beta}_{\alpha}$ (in field theory, the 
summation over like indices implies a spatial integration, also), so 
that the secondary reality condition is satisfied when the primary one 
is.  However, this relation is of no value in implementing the 
tertiary condition.  Instead, if we had
\begin{equation}
    \{f_{\alpha},\,H_{\rm c}\}
    = \eta^{\beta}_{\alpha}\,f_{\beta}
        \label{eta} 
\end{equation}
for any real matrix $\eta^{\beta}_{\alpha}$ such that 
\[  \{\eta^{\beta}_{\alpha},\,H_{\rm c}\} =0 \ ,
\] 
then indeed the stabilization algorithm would have been over.  Of 
course this is only a sufficient condition.

In a more realistic case we would use $H_{\rm D}$, which is in general 
time dependent.  Considering how we arrived at (\ref{recurr3}), which 
plays, for the reality conditions, the role analogous to 
(\ref{conset}) for Dirac constraints, it is easy to get an analog for 
(\ref{recurr2}).  In fact we can use here all the results obtained 
from the Dirac analysis, in particular the determination in phase 
space of some of the Lagrange multipliers.  This means that we can 
start with a first class ($fc$) Hamiltonian
\[   H^{fc}_{\rm D} 
    = H_{\rm c} + \sum_{\mu=1}^{n_{1}}\lambda^{\mu}_{\rm c} \phi_{\mu} 
        + \sum_{\mu=n_{1}}^{n}\lambda^{\mu} \phi_{\mu}\ ,
\]
where we have assumed for simplicity that the first $n_{1}$ Lagrange 
multipliers are the ones that get determined as functions 
$\lambda^{\mu}_{\rm c}$ in phase space through the Dirac stabilization 
algorithm.  In this general case the reality conditions may lead to a 
further reduction of the gauge freedom present in $H^{fc}_{\rm D}$, 
that is, to a partial determination of the remaining Lagrange 
multipliers---for instance: their real or imaginary parts.  This is 
what will happen with the triad reality conditions for the Ashtekar 
formulation, to be analyzed in Section \ref{sec:real}.

It is obvious that nothing in this Section depends on the theory being 
formulated in phase space.  Indeed, we could replace 
$\{-,\,H_{\rm c}\}+\lambda^{\mu}\{-,\,\phi_{\mu}\}$ everywhere by 
${\bf X}+\lambda^{\mu}{\bf Y}_{\mu}$, with ${\bf X}$ and 
${\bf{}Y}_{\mu}$ being vector fields in some given space (for instance 
configuration-velocity space).


\section{The Ashtekar Lagrangian}
\label{sec:lag}

One way to present the Ashtekar Lagrangian density is 
\cite{samuel/87,jacobson/smolin/87,jacobson/smolin/88}
\begin{equation}
    {\cal L_A} 
    =  {}^{4}\!F^{IJ}_{\tau\sigma}[{}^{4}\!A]
        E^{\mu}_{I}E^{\nu}_{J} \sqrt{|g|} \ ;
                \label{ast-l}
\end{equation}
where $g$ is the determinant of the spacetime
metric; $E^{\mu}_{I}$ are the tetrad components, $\mu$ being a spacetime
index and I an internal index; and ${}^{4}\!F^{IJ}_{\tau\sigma}$ is
the curvature tensor associated with the Ashtekar connection
${}^{4}\!A^{IJ}_{\mu}$.  We use the standard definitions of these
quantities \cite{rovelli/91}, and we do not repeat these definitions
here, because we will be working in a $3+1$ decomposition and will
give specific definitions of our variables below.

$\cal L_{A}$ is interpreted in a Palatini-like formalism: The 
components of the self-dual complex connection are taken to be 
independent variables.  Their equations of motion determine them in 
terms of the other variables (and their derivatives).  This 
determination is similar to the determination of the Christoffel 
coefficents in the Einstein-Palatini version of general relativity 
(see \cite{peldan/94} for a good review of actions for gravity).  
Variables having this property of being determined by their own 
equations of motion are usually called auxiliary variables.  When this 
dynamical determination of the Ashtekar connection is substituted into 
the Lagrangian we get the standard Ashtekar Lagrangian, which is 
equivalent to the Einstein-Hilbert Lagrangian.

We are interested in the canonical description (in phase space).  
Therefore we will write the action in a $3+1$ decomposition of the 
variables.  The contravariant spacetime metric is written in terms of 
the lapse function $N$ and shift vector $N^{a}$, and a triad of 
orthonormal vectors $T^a_i$ ($a,b$ are spatial indices; $i,j$ are 
internal indices, raised or lowered with $\delta_{ij}$, so that 
repeated internal indices imply a sum even if both are raised or 
lowered):
\begin{equation}
    g^{\mu\nu} 
    = \left(\begin{array}{cc}
        -N^{-2}\quad      & N^{-2} N^a \\
        N^{-2} N^b \quad  & T^a_i T^{b}_{i}-N^{-2} N^a N^b
    \end{array}\right)\ .
        \label{themetric} 
\end{equation}
The triad vectors and the (unit) normal vector to the constant-time 
hypersurfaces
\[  n^\mu = (N^{-1},-N^{-1}N^a)
\] 
constitute an orthonormal tetrad.  

We represent the components of the orthonormal spatial one-forms by 
$t^i_a$, so that the covariant three-metric is given by
\[  g_{ab} = t^i_a t^{i}_{b}\ .
\]  
It turns out to be convenient to take one set of canonical variables 
to be the triad vectors multiplied by the square root of the 
determinant of the three-metric.  As has now become conventional, we 
represent densities of arbitrary positive weight under spatial 
diffeomorphisms by an appropriate number of tildes over the symbol.  
For negative weights we place the tilde(s) below the symbol.  Hence we 
define, for
\[  t := \sqrt{\det(g_{ab})} = \det(t^i_a)\ ,
\]
the densitized triad as
\begin{equation}
    \OT{T}^a_i := t \, T^a_i \ .
        \label{ttilde}
\end{equation}

In the Ashtekar approach the connection is self-dual.  An 
antisymmetric tensor, whose components in an orthonormal tetrad are 
$F_{IJ}$, is self-dual if
\[  i F_{IJ} = {1\over2}\epsilon_{IJKL}F^{KL} \ ,
\]
where $\epsilon_{IJKL}$ is the four-dimensional Levi-Civita symbol 
defined by $\epsilon_{0123}=-1$.  Because of self-duality, the 
four-connection ${}^{4}\!A^{IJ}_{\mu}$ in (\ref{ast-l}) is determined 
by the independent components
\[  A^{i}_{\mu} 
    := {1\over2} \epsilon^{ijk}A_{\mu}^{jk}\ ,
\]
$\epsilon^{ijk}$ being the Levi-Civita symbol.  In the $3+1$ 
decomposition the Ashtekar Lagrangian becomes ( $\dot{}$ is 
$\partial/\partial x^0$, and we will also use a subscript comma for 
partial derivatives)
\begin{eqnarray}
    {\cal L_{A}} 
    &=&  i \dot{\OT T^a_{i}} A_a^i
        - i A_0^i{\cal D}_a \OT T^a_{i}
        + i N^a \OT T^b_{i} F_{ab}^{i}
            \nonumber \\
    && + {1\over2}{\UT N}\OT T^a_{i} \OT T^b_{j} F_{ab}^{ij}\ ;
                \label{thelag}        
\end{eqnarray}
where $F_{ab}^{jk}=:\epsilon^{ijk}F_{ab}^{i}$ is the three 
dimensional Riemann tensor associated with the Ashtekar connection,
\[  F_{ab}^{i} 
    := A_{b,a}^{i} - A_{a,b}^{i} 
        - \epsilon^{ijk}A_{a}^{j} A_{b}^{k}\ ;
\]
and where the covariant derivative ${\cal D}_{b}$ is defined using the 
Ashtekar connection:  Its action on the densitized triad is, for 
example,
\begin{equation}
    {\cal D}_{b} \OT T^{a}_{i} 
    = {}^{3}\nabla_{b} \OT T^{a}_{i}
                   +\epsilon^{ijk} \OT T^{a}_{j}A^{k}_{b}\ ,
               \label{covderAsh}
\end{equation}
${}^{3}\nabla_{b}$ being the covariant derivative based on the 
3-metric $g_{ab}$.  It is convenient to take the densitized lapse 
$\UT{N}$ as an independent variable, but for convenience, some 
equations will be written in terms of $N$ itself; likewise it will 
prove convenient to use both densitized and undensitized variables in 
some of our results.

Two observations should be made at this point. 

First: From the fact that ${\cal L}$ in (\ref{thelag}) does not depend 
on the velocities $\UT{{\dot N}}$, $\dot{N}^{a}$, $\dot A_{0}^{i}$, we 
can conclude (details are given in \cite{pons/salisbury/shepley/97}) 
that the necessary and sufficient condition for a function $f$ in 
configuration-velocity space $TQ$ to be projectable to phase space 
$T^{*}\!Q$ is that $f$ does not depend on these velocities.

Second: The fact that the independent components of the Ashtekar 
connection play the role of auxiliary variables tells us that their 
equations of motion give
\begin{equation}
    A^{i}_\mu - \Omega_\mu^{i} - i\Omega_\mu^{0i} 
    = 0 \ ,
        \label{ast}
\end{equation}
where $\Omega_\mu^{i}:={1\over2}\epsilon^{ijk}\Omega_\mu^{jk}$ and 
$\Omega_\mu^{0i}$ are the components of the spin connection, that is, 
the Ricci rotation coefficients.  In particular, $\Omega_a^{ij}$ are 
the three-dimensional Ricci rotation coefficients formed from the 
triad, so that
\begin{mathletters}\label{omeg}
\begin{equation}
    \Omega_a^{i} 
    := {1\over2}\epsilon^{ijk} \, t^{j}_{b}
        \bigr(T^{b}_{k,a} + {}^{3}\Gamma^b_{ca}T^{c}_{k}\bigr) 
    =: \omega^{i}_{a} \ ,
            \label{omeg-aij}
\end{equation}
with ${}^{3}\Gamma^b_{ca}$ being the Christoffel symbols.  For future 
use, we define the covariant derivative using the three-dimensional 
Ricci coefficients, which applied to $\OT T^{a}_{i}$ gives zero:
\begin{equation}
    D_{b} \OT T^{a}_{i} 
        = {}^{3}\nabla_{b} \OT T^{a}_{i}
           + \epsilon^{ijk} \OT T^{a}_{j}\omega^{k}_{b} 
        = 0 \ .
            \label{DDDDD}
\end{equation}
Notice that when (\ref{ast}) holds,
\[  ({\cal D}_{a}- D_{a}) \OT T^{b}_{i}
    = i \epsilon^{ijk} \OT T^{b}_{j} \Omega^{0k}_{a} \ .
\]
The other components of the spin connection involve time derivatives:
\begin{eqnarray}
    \Omega_0^{i} 
    &:=& {1\over2}\epsilon^{ijk}\bigr(\dot t_a^{j}T^{a}_{k}
        + N^b_{,a}t^{k}_{b}T^{a}_{j} 
                \nonumber \\
    &&\quad + t^{k}_{b,a}T^{a}_{j}N^b
         +t^\ell_{b,a}N^c t^\ell_c T^{a}_{j}T^{b}_{k}\bigr)\ , 
                \label{omeg-0ij} \\
    \Omega_a^{0i}
    &:=& T^{b}_{i}K_{ab}\ , 
                \label{omeg-a0i} \\
    \Omega_0^{0i}
    &:=& T^{a}_{i} N_{,a} + N^a T^{b}_{i}K_{ab}\ , 
                \label{omeg-00i}
\end{eqnarray}
\end{mathletters}%
where $K_{ab}$ is the extrinsic curvature, defined as  
\[  K_{ab}
    := {1\over 2N}(\dot g_{ab} - N^{c} g_{ab,c} 
        - g_{ca} N^{c}_{,b} - g_{bc} N^{c}_{,a}) \ .
\]
Equations (\ref{ast}, \ref{omeg}) will be useful when we consider the 
reality conditions and in determining variations of $A^{i}_{0}$.  Now 
we will continue with the canonical version of the theory.


\section{The canonical Hamiltonian approach}
\label{sec:can}

The Legendre map
\[  {\cal F}\!L : TQ \longrightarrow T^{*}\!Q
\]
from configuration-velocity (tangent) space to phase space is defined 
by
\[  {\cal F}\!L(q,\dot q) 
        = (q, p = \hat p := {\partial L\over\partial\dot q})\ ;
\] 
we working locally, with $q,\dot q$ being coordinates in 
configuration-velocity space and $q,p$ being coordinates in phase 
space, as is conventional.

Our configuration variables and their conjugate canonical momenta are 
as follows: 
\begin{eqnarray*}
    A_\mu^i\ && ({\rm canonical\ momenta:}\ \OT\pi^\mu_i) \ , \\
    \OT T{}^a_{i}\ && ({\rm canonical\ momenta:}\ P_a^i) \ , \\
    {\UT N}\ &&({\rm canonical\ momentum:}\ \OTT P) \ , \\
    N^a\ && ({\rm canonical\ momenta:}\ \OT P_{a}) \ .
\end{eqnarray*}
The primary constraints, consequences of the Lagrangian definition of 
the momenta, are:
\begin{eqnarray*}
    \OTT P &=& 0  \ , \\
    \OT P_{a} &=& 0  \ , \\
    \OT\pi^\mu_i &=& 0  \ , \\
    P_a^i - i A_a^i &=& 0  \ . 
\end{eqnarray*}

The canonical Hamiltonian $H_c$ is defined as a function in phase 
space such that its pullback to tangent space under the Legendre map 
is the Lagrangian energy $E_{L}$ from (\ref{eq:lageng}), that is, 
$E_{L}={\cal F}\!L^{*}(H_{\rm c})$.  $H_{\rm c}$ is uniquely defined 
up to primary constraints.  We take
\begin{eqnarray}
    H_c 
    &=& \int d^{3}x \bigr(i A_0^i{\cal D}_a{\OT T{}^a_{i}}
                            - i N^a \OT T{}^b_{i}F_{a b}^{i}
            \nonumber \\
    &&\qquad\qquad -{1\over2}{\UT N} \OT T{}^a_{i}
                            \OT T{}^b_{j} F_{ab}^{ij}\bigr) \ .
            \label{hc}
\end{eqnarray}
The constraints $P_a^i-iA_a^i=0$ and $\OT\pi^a_i=0$ are second class 
in the sense of Dirac and can be readily disposed of; in the process, 
we eliminate the conjugate variables $A_a^i$ and $\OT \pi^a_i$.  The 
recipe is to put $A_a^i=-iP_a^i$ and $\OT \pi^a_i=0$ everywhere in the 
Hamiltonian.  In fact, we don't even need to substitute $-iP_a^i$ for 
$A_a^i$: Since $P_a^i$ was not present in $H_c$, we can just take 
$iA_a^i$ to be the momentum variable canonically conjugate to ${\OT 
T{}^a_{i}}$.  The rest of the variables are pairs of conjugate 
variables whose Dirac brackets coincide with the Poisson brackets.

We have achieved a canonical Hamiltonian $H_c$, and a number of 
canonical variables with Poisson brackets (actually Dirac brackets),
\begin{mathletters}\label{PB}
\begin{eqnarray}
    \{{\UT N}, \, \OTT P' \} &=& \delta^{3}(x-x') \ ,
        \label{PB.a} \\
    \{N^a, \, \OT P'_{b}\} &=& \delta^a_b\delta^{3}(x-x') \ ,
        \label{PB.b} \\
    \{{\OT T{}^a_{i}}, \,A'{}_b^j\} 
        &=& -i \, \delta^a_b\delta^j_i\delta^{3}(x-x') \ ,
            \label{PB.c} \\
    \{A_0^i, \, \OT\pi'^0_j \} &=& \delta^i_j\delta^{3}(x-x') \ .
        \label{PB.d}
\end{eqnarray}
\end{mathletters}%

The Dirac Hamiltonian, which governs the time evolution of the system, 
is constructed by adding to $H_{\rm c}$ the primary constraints 
multiplied by arbitrary functions:
\begin{equation}
    H_D = H_c + \int d^{3}x 
        \left(\UT\lambda \OTT P + \lambda^a \OT P_{a} 
            + \lambda^i \OT\pi^0_i\right) \ .
        \label{hd}
\end{equation}
The second class primary constraints having been already eliminated, 
all the remaining primary constraints are first class.  

The equations of motion derived from $H_{\rm D}$ for ${\OT T}{}^a_{i}$
and $A^i_{a}$ are
\begin{mathletters}\label{eq-ta}
\begin{eqnarray}
    \dot{\OT T}{}^a_{i}
    &=& \epsilon^{ijk} \OT T^a_{k} A_0^j 
        + 2 {\cal D}_b (N_{\strut}^{[b} \OT T{}^{a]}_{i}) 
            \nonumber \\
    &&    -i\epsilon^{ijk} {\cal D}_b ({\UT N} 
            \OT T{}^{b}_{j} \OT T{}^{a}_{k}) \ ,
            \label{eq-t} \\
    \dot{ A}^i_{a} 
    &=& {\cal D}_a A^i_{0} + N^b F_{ba}^{i} 
	    - i{\UT N} \OT T^b_{j} F_{ab}^{ij} \ .
            \label{eq-a}
\end{eqnarray}
\end{mathletters}%
The equations obtained from the stabilization of the primary first 
class constraints yield the three secondary constraints
\begin{mathletters}\label{second}
\begin{eqnarray}
    \OTT{\cal H}_0 & := & - {1\over2}\OT T^a_{i}
        \OT T^b_{j} F_{a b}^{i j}= 0\ ,
            \label{seconda}\\
    \OT{\cal H}_a & := & -i \OT T^b_{i}F_{a b}^{i} = 0\ , 
            \label{secondb}\\
    \OT{\cal H}_i & := & -i{\cal D}_a \OT T^a_{i} = 0\ . 
            \label{secondc}
\end{eqnarray}
\end{mathletters}%
The canonical Hamiltonian written in terms of these constraints is
\begin{equation}
    H_{\rm c}
    =  \int d^{3}x \, (-A^{i}_{0}\OT{\cal H}_{i}
      + N^{a}\OT{\cal H}_{a} + \UT N \OTT{\cal H}_{0}) \ .
          \label{cancan}
\end{equation}
Finally, the equations for the rest of the variables, $\UT N$, 
$N^{a}$, $A^{i}_{0}$, are
\[  \UT{{\dot N}} = \UT\lambda\ , \ 
    \dot{N}^{a} = \lambda^{a}\ , \ 
    \dot{A}^{i}_{0} = \lambda^{i} \ .
\] 
They inform us that these variables are arbitrary---gauge---variables.
The secondary constraints (\ref{second}) are all first class (their 
algebra will be displayed in Section \ref{sec:gen}).  No more 
constraints appear.

Let us observe that the Lagrangian equations of motion for ${\OT 
T}{}^a_{i}$ and ${A}^i_{a}$ are the same as the Hamiltonian equations 
of motion.  The constraints (\ref{second}) appear in 
configuration-velocity space as the Lagrangian equations of motion for 
the variables $\UT N$, $N^{a}$, and $A^{i}_{0}$.  There are no
equations for the time derivatives of these variables, indicating 
that they are gauge 
variables.  Also, observe that equations (\ref{ast}) have the same 
contents as (\ref{eq-t}) and (\ref{secondc}).

Now we are ready to apply our stabilization procedure for the reality
conditions to Ashtekar's version of 
canonical gravity.


\section{The reality conditions for Ashtekar canonical gravity}
\label{sec:real}

\subsection{The metric reality conditions}
\label{subsec:real.metric}

At the very least, the metric tensor should be real: the primary 
metric reality conditions are
\begin{mathletters}\label{pmrc}
    \begin{eqnarray}
        \Im\, \UT N &=& 0 \ ,
                \label{pmrc.N}\\
        \Im\, N^{a} &=& 0 \ ,
                \label{pmrc.Na}\\
        \Im\, \OTT e{}^{ab} &=& 0 \ ,
                \label{pmrc.g}
\end{eqnarray}
\end{mathletters}%
where $\OTT e{}^{ab} = \OT{T}{}^{a}_{i} \OT{T}{}^{b}_{i}$.  It is
clear that, according to (\ref{hd}), (\ref{pmrc.N}) and
(\ref{pmrc.Na}) fix the arbitrary functions $\UT\lambda$ and
$\lambda^{a}$ to be real.  These equations do not have any further
consequence.  Requirement (\ref{pmrc.g}) is equivalent to
$\Im{g}_{ab}=0$.  Notice that these reality conditions will also 
preserve the Lorentzian signature of the metric (presuming that $N$ 
and $\det(T^{a}_{i})$ remain nonzero).

Before applying our method of stabilization, let us recall the last
result in Section \ref{sec:lag}: The components of the Ashtekar
connect ion are auxiliary variables for the Lagrangian (\ref{thelag}). 
Recalling the definitions (\ref{omeg-a0i}), we can write a portion of
the equations of motion (\ref{ast}) as
\begin{equation}
    A_a^{i} - \omega_a^{i} = +iT^b_i K_{ba} \ .
\end{equation}
Thus, if we define the quantities $M_{ab}$ as
\begin{equation}
    M_{ab} := -i t^{i}_{a}(A_b^{i} - \omega_b^{i}) \ ,
        \label{them}
\end{equation}
then this portion of the equations of motion becomes
\begin{equation}
    K_{ab} = M_{ab}\ .
        \label{k=g}
\end{equation}
$K_{ab}$ is a functional of the three-metric that is real and 
symmetric.  Thus we find here a requirement that $M_{ab}$ must be real 
and symmetric.  The symmetry is already guaranteed by the constraint 
(\ref{secondc}).  That $M_{ab}$ must be real is in fact the content of 
the secondary reality conditions, $\Im\{{g}_{ab},\,H_{\rm c}\} =0$, as 
we shall now prove.

The equations of motion for $g_{ab}$ are hidden in (\ref{k=g}),
\begin{equation}
    \dot{g}_{ab} = \{{g}_{ab},\, H_{\rm D}\} 
        = \{{g}_{ab},\, H_{\rm c}\} 
        = 2 N M_{ab} 
            + {\cal L}_{\vec N}({g}_{ab}) \ ,
\label{secmrc}
\end{equation}
where ${\cal L}_{\vec N}$ is the Lie derivative with respect to the 
vector field $N^{c}\partial_{c}$.  From the first term in 
(\ref{secmrc}) we extract the secondary reality conditions
\begin{equation}
    \Im M_{ab} =0 \ ,
        \label{2ndreal}
\end{equation}
as was expected.  

The last term in (\ref{secmrc}) is a combination of the type 
$\eta^{\nu}_{\mu} f_{\mu}$, as discussed in (\ref{eta}), with 
\begin{eqnarray}  
    \eta(x,x')^{cd}_{ab} &=&  
    N^{e}\delta^{3}_{,e}(x-x')\delta^{c}_{a}\delta^{d}_{b}
            \nonumber \\
    &&\quad    + N^{c}_{,a}\delta^{3}(x-x')\delta^{d}_{b}
        + N^{d}_{,b}\delta^{3}(x-x')\delta^{c}_{a} 
            \nonumber \ .
\end{eqnarray}
We had mentioned that the stabilization procedure simplifies when 
$\{\eta,\,H_{\rm D}\}$ vanishes; a similar simplification occurs when, 
as here, $\{\eta,\,H_{\rm D}\}$ is not zero but a harmless combination 
of the $\lambda^{a}$ (which are real).  Thanks to this fact, and 
applying a similar argument to show the irrelevance of the factor $N$ 
before $M_{ab}$ in (\ref{secmrc}), we are ready to consider the 
tertiary reality conditions.

Since $\{M_{ab},\,H_{\rm D}\}= \{M_{ab},\,H_{\rm c}\}$, the tertiary 
reality conditions are
\begin{equation}
    \Im\{M_{ab},\,H_{\rm c}\} =0 \ .
        \label{termrc}
\end{equation}
The computation of (\ref{termrc}) is a bit involved.  It is useful to 
start by writing the canonical Hamiltonian (\ref{hc}) as a sum of 
three terms that clearly preserve the reality of a real triad.  This 
way we will also gain information on the structure of the Hamiltonian; 
this information is useful whether we consider the metric or the triad 
reality conditions.

The term $N^{a}\OT{\cal H}_{a}$ (we have used the definition 
(\ref{secondb})) in $H_{\rm c}$ produces a time evolution of the triad 
that makes it acquire an imaginary part.  This part can be eliminated 
by a rotation generated by $\OT{\cal H}_{i}$.  This way we obtain a 
unique linear combination of $\OT{\cal H}_{a}$ and $\OT{\cal H}_{i}$ 
that preserves the reality of a real triad.  We are led to define
\begin{equation}
    \OT{\cal G}_{a} := \OT{\cal H}_{a} - A^{i}_{a}\OT{\cal H}_{i} \ .
         \label{ha}
\end{equation}
Then $H_{\rm c}$ is written as 
\begin{equation}
    H_{\rm c}
    = \int d^{3}x \left(-(A^{i}_{0} - N^{a} A^{i}_{a})\OT{\cal H}_{i}
          +N^{a}\OT{\cal G}_{a} + {\UT N}\OTT{\cal H}_{0}\right) \ . 
              \label{hhhhh}
\end{equation} 

The rotations generated by the first term in (\ref{hhhhh}), the 
integrand of which is equal to $-NA^{i}_{\mu}n^{\mu}\OT{\cal H}_{i}$, 
are not real in general.  But note that according to the equations of 
motion (\ref{ast})
\begin{equation}
    NA^{i}_{\mu}n^{\mu} - i T^{b}_{i}N_{,b} 
    = N\Omega^{i}_{\mu}n^{\mu} \ ,
        \label{obstacle}
\end{equation}
where we have used definitions (\ref{omeg-a0i}) and (\ref{omeg-00i}).  
Since $\Omega^{i}_{\mu}$ will be real if the triad reality conditions 
hold, it is useful to rewrite $H_{\rm c}$ as
\begin{eqnarray}
    H_{\rm c}
    &=& \int d^{3}x \Big(-(A^{i}_{0} - N^{a} A^{i}_{a} 
                 - i \OT T^{b}_{i}{\cal D}_{b}\UT N)\OT{\cal H}_{i}         
            +(N^{a}\OT{\cal G}_{a})
                    \nonumber \\
    &&\qquad\quad + \bigr({\UT N}\OTT{\cal H}_{0} 
      - i\OT T^{b}_{i}({\cal D}_{b}\UT N)\OT{\cal H}_{i} \bigr)\Big)\ .
                \label{newhc}  
\end{eqnarray}
Let us display the action of these three terms of $H_{\rm c}$ on 
$t^{i}_{a}$ and $A^{i}_{a}$ (since we are computing (\ref{termrc}), 
recall that $M_{ab} :=-i t^{i}_{a}(A_b^{i} - \omega_b^{i})$).

The first term in (\ref{newhc}) is of the type 
\begin{equation}
    \int d^{3}x \, B^{i} \OT{\cal H}_{i} \ ,
\label{newhc1}  
\end{equation}
with $B^{i}$ complex.  It generates $SO(3,C)$ rotations ($R$) of the 
triad vectors, $\delta\tau$ being an infinitesimal parameter,
\[  \delta_{R}[B\delta\tau] t^{i}_{a} 
    = - \epsilon^{ijk} B^{j} t^{k}_{a}\delta\tau \ ,
\]
and for the connection components, 
\[
    \delta_{R}[B\delta\tau] A^{i}_{a} 
    = - {\cal D}_{a}B^{i} \delta\tau \ ,
\]
that is, the Yang-Mills-like gauge transformation.  The variations of 
the Ricci rotation coefficients are computed from the variations of 
the triad vectors, the results being
\[  \delta_{R}[B\delta\tau] \omega^{i}_{a} 
    = - D_{a}B^{i} \delta\tau \ .
\]
where $D_{a}$ stands for the covariant derivative associated 
with the spin connection $\omega^{i}_{a}$.    

The second term in (\ref{newhc}) is 
\begin{equation}
    \int d^{3}x \, N^{a} \OT{\cal G}_{a} \ .
        \label{newhc2}  
\end{equation}
It generates standard spatial (three-space) diffeomorphisms ($D$), 
that is,
\begin{eqnarray}
    \delta_{D}[\vec N\delta\tau] t^{i}_{a}
        &=& (N^{b}t^{i}_{a,b} + t^{i}_{b} N^{b}_{,a})\delta\tau \ ,
            \nonumber\\
    \delta_{D}[\vec N\delta\tau] A^{i}_{a}
    &=& (N^{b}A^{i}_{a,b} + A^{i}_{b} N^{b}_{,a})\delta\tau \ .
            \nonumber
\end{eqnarray}

The third term in (\ref{newhc}) generates a perpendicular
diffeomorphism (that is, perpendicular to the
constant-time hypersurfaces) plus a gauge rotation with descriptor
\[  t \UT N A^i_\mu n^\mu -i \OT T^{bi}{\cal D}_b \UT N \ ,
\]
as we will show in Section \ref{sec:proj}.  Thus in the real triad 
sector it does generate real variations.  These variations (which we 
call $\delta_{S'}$ to distinguish them from the variations 
$\delta_{S}$ generated by $\OTT{\cal H}^{0}$) are in fact identical to 
the variations generated by the scalar generator in the real triad 
formalism \cite{pons/salisbury/shepley/99b}, although here we apply 
them even if the triad is not real.  The resulting variation is
\begin{equation}
    \delta_{S'}[\UT N\delta\tau]  \OT T^{a}_{i}
    = - i  \epsilon^{ijk}{\cal D}_{b}
        (\OT T^{b}_{j}\OT T^{a}_{k}) {{\UT N}} 
             \delta\tau  \ .
    \label{projdelta}
\end{equation}
The corresponding variation of $t^{i}_{a}$ is
\[  \delta_{S'}[\UT N\delta\tau] t^{i}_{a} 
    = t \UT N M^{b}_{a} t^{i}_{b} \delta\tau \ ,
\]
where $M^{b}_{a} = e^{bc}M_{ca}$, with  
\[  e^{ac}g_{cb}= \delta^a_b \ .
\]
When operating on $A^{i}_{a}$ this transformation is, on the 
constraint hypersurfaces, 
\[  \delta_{S'}[\UT N\delta\tau] A^{i}_{a} 
    = -i\left(\UT N \OT T^{b}_{j} F^{ij}_{ab} 
        - {\cal D}_{a}(\OT T^{b}_{i} {\cal D}_{b}\UT N)
            \right)\delta\tau \ .
\]  
The variations of the Ricci rotation coefficients are computed from 
the variations of the triad vectors,
\begin{equation}
    \delta_{S'}[\UT N\delta\tau] \omega^{i}_{a} 
    = \epsilon^{ijk} \OT T^{b}_{k} T^{c}_{j}
        {\cal D}_{b}(\UT N M_{ac}) \delta\tau \ .
            \label{ddOHM}
\end{equation}

Now we can compute $\{M_{ab},\,H_{\rm c}\}$.  The result is
\begin{eqnarray}
    \{M_{ab},\,H_{\rm c}\} 
    &=& N(- {}^{3}\!R_{ab} + M^{c}_{c} M_{ab})
            \nonumber \\ 
    &&+ {\cal L}_{\vec N}{M}_{ab} 
        + {\cal D}_a{\cal D}_b N \ ,
\label{longcomp}
\end{eqnarray}
where the symmetry of $M_{ab}$ (guaranteed by the constraint 
(\ref{secondc})) has been used, and ${}^{3}\!R_{ab}$ is the three 
dimensional Ricci tensor.

Therefore the tertiary reality conditions (\ref{termrc}) are 
automatically satisfied, for all terms on the right side of 
(\ref{longcomp}) are real by way of the primary and secondary reality 
conditions.

Also, we have more information: The first term on the right side of 
(\ref{longcomp}) is of the type (\ref{eta}) and will not give further 
consequences in subsequent levels of stabilization.  The same is true 
for all the other terms, though they are not exactly of the type 
(\ref{eta}).  For instance, consider the term $N_{,ab}$ in the last 
term of (\ref{longcomp}).  In stabilizing this term, notice that 
$\{N_{,ab},\,H_{\rm D}\} = \lambda_{,ab}$ which is already real.  The 
next step, $\{\lambda_{,ab},\,H_{\rm D}\}$, gives exactly zero.

Summing up, from the form of the right side of (\ref{longcomp}) we 
conclude that the metric reality conditions have been fully satisfied.  
The algorithmic procedure devised in the previous Section has 
terminated.


\subsection{The triad reality conditions}
\label{subsec:real.triad}

The primary triad reality conditions are
\begin{mathletters}\label{ptrc}
    \begin{eqnarray}
        \Im \UT N &=& 0 
            \label{ptrca}\\
        \Im N^{a} &=& 0 
            \label{ptrcb}\\
        \Im\, \OT{T}{}^{a}_{i} &=& 0 \ .
            \label{ptrcc}
\end{eqnarray}
\end{mathletters}%
As before, (\ref{pmrc.N}) and (\ref{pmrc.Na}) fix the arbitrary 
functions $\UT\lambda$ and $\lambda^{a}$ to be real.  They do not have 
any further consequence.

The secondary reality conditions are 
$\Im\{\OT{T}{}^{a}_{i},\, H_{\rm c}\} =0$:
\begin{eqnarray}
    \Im\{\OT{T}{}^{a}_{i},\, H_{\rm c}\} 
    &=& \epsilon^{ijk} \OT T^{a}_{k}\Im A^{j}_{0} 
               \nonumber\\
    && + 2 \epsilon^{ijk} N^{[b}\OT T^{a]}_{j} \Im A^{k}_{b}
       - \epsilon^{bac} N_{,b} t^{i}_{c} 
               \nonumber\\
    && - \epsilon^{bac}\epsilon^{ijk} N t^{j}_{c} 
            (\Re A^{k}_{b} - \omega^{k}_{b}) \ .
                   \label{stabt}
\end{eqnarray}
Using the primary triad reality condition (\ref{ptrcc}), we can write 
\[  \Re A^{k}_{b} - \omega^{k}_{b} =+ T^{d}_{k}\Im M_{bd} \ .
\]
Computing 
$T^{b}_{i}\Im\{\OT{T}{}^{a}_{i},\, H_{\rm c}\}+(a\leftrightarrow b)$,
we get 
\begin{equation}
    \Im M_{ab} = 0 \ ,
         \label{sectrc1}
\end{equation}
where the constraint (\ref{secondc}) has been used.  These secondary 
reality conditions (\ref{sectrc1}) were expected from the calculations 
of the metric reality condition case.  The remaining terms of 
(\ref{stabt}) give the rest of the secondary triad reality 
conditions,
\begin{equation}
    \Im\left(A^{i}_{0} - N^{a}A^{i}_{a} 
        - i \OT T^{b}_{i}{\cal D}_{b}\UT N \right) =0 \ . 
            \label{sectrc2}
\end{equation}
Notice that the object in (\ref{sectrc2}) which is required to be real 
is the coefficient of $\OT{\cal H}_{i}$ in $H_{\rm c}$ in 
(\ref{newhc}).

We need not worry about the stabilization of (\ref{sectrc1}) because 
this issue has been already addressed in the study of the metric 
reality conditions.  We do have to be concerned with the stabilization 
of (\ref{sectrc2}).  The tertiary triad reality conditions read
\begin{equation}
    \Im\{(A^{i}_{0} - N^{a}A^{i}_{a} 
        - i \OT T^{b}_{i} {\cal D}_{b}N),\ H_{\rm D}\} 
    =0 \ .  
                \label{sectrc3}
\end{equation}
They determine the imaginary part of $\lambda^{i}$ in (\ref{hd}),
\begin{eqnarray}
    \lambda^{i} 
    &=& \lambda^{i}_{0}+\lambda^{a} A^{i}_{a} 
        + i \OT T{}^a_{i} {\cal D}_{a} \UT\lambda 
        + N^{a}\{A^{i}_{a},\,H_{\rm c}\} 
                \nonumber \\
        &&\quad + i\{\OT T{}^{ai}, \, H_{\rm c}\}
                {\cal D}_{a}{\UT N} \ ,
                    \nonumber
\end{eqnarray}
where $\lambda^{i}_{0}$ is a real arbitrary function.  Notice that we 
have reduced the gauge freedom of rotations of the triad vectors from 
$SO(3,C)$ to $SO(3,R)$.

With this determination, the Dirac Hamiltonian becomes
\begin{eqnarray} 
    H_{\rm D}' 
    &=& H_{\rm c}+ \int d^{3}x
        \Big(\lambda^{a} A^{i}_{a} 
              + i \OT T{}^a_{i}{\cal D}_{a} \UT\lambda 
              + N^{a} \{A^{i}_{a},\, H_{\rm c}\} 
                  \nonumber\\
      &&\qquad\qquad + i\{\OT T{}^{a}_{i}, \, H_{\rm c}\}
              {\cal D}_{a}{\UT N} \Big) \OT\pi^0_i 
                  \nonumber\\
     && + \UT\lambda \OTT P + \lambda^a \OT P_{a} 
         + \lambda^i_{0} \OT\pi^0_i \ ,
\end{eqnarray} 
with $\UT\lambda$, $\lambda^a$, and $\lambda^i_{0}$ all real arbitrary 
functions. 

$H_{\rm D}'$ is now used for time evolution.  The next reality 
condition is
\begin{equation}
    \Im\{\{(A^{i}_{0} - N^{a}A^{i}_{a} 
        - iT^{b}_{i}N_{,b}), \, H_{\rm D}'\},\, H_{\rm D}'\} 
    =0 \ ,  
            \label{sectrc4}
\end{equation}
which is trivially satisfied:  Since now 
\[  \{(A^{i}_{0} - N^{a}A^{i}_{a} - iT^{b}_{i}N_{,b}),\,H_{\rm D}'\}
    = \lambda^{i}_{0} \ ,
\] 
we have the stronger result
\begin{equation}
    \{\{(A^{i}_{0} - N^{a}A^{i}_{a} - iT^{b}_{i}N_{,b}), \, 
        H_{\rm D}'\},\,H_{\rm D}'\} 
    =0 \ ,  
\end{equation}
which guarantees that no further reality conditions will arise.


\section{Projectability of gauge symmetries}
\label{sec:proj}

In this Section we will realize the full gauge group in phase space, 
including transformations based on spacetime diffeomorphisms and triad 
rotations.  Two tasks are involved in this goal.  The first one is to 
make the infinitesimal gauge transformations in configuration-velocity 
space projectable to phase space.  From our previous experience with 
conventional general relativity \cite{pons/salisbury/shepley/97}, 
Einstein-Yang-Mills theory \cite{pons/salisbury/shepley/99a}, and real 
triad theory \cite{pons/salisbury/shepley/99b}, we know that the 
arbitrary functions in the infinitesimal spacetime diffeomorphisms 
must depend in an explicit way on the lapse and shift functions.  This 
was sufficient in the case of general relativity, but in the latter 
two cases a second step was required: We needed to add a gauge 
rotation.  We expect something similar to occur with the Ashtekar 
formulation.

The second task is to construct the generators of the gauge group in 
phase space and to check that the transformations they generate do 
indeed coincide with the projectable transformations in 
configuration-velocity space.  Notice that now there is a consistency 
condition to be met which wasn't needed in our previous work: We must 
require that the gauge group preserve the reality conditions.

We have already calculated \cite{pons/salisbury/shepley/99b} the 
projectable variations of the configuration variables $\UT N$ and 
$N^{a}$ under diffeomorphisms with
\[  x^{\mu}\rightarrow
    x^{\mu}-\delta^{\mu}_{a}\xi^{a} - n^{\mu} \xi^{0}\ ,
\]
where the $\xi^{\mu}$ are arbitrary functions.  As in all the theories 
considered previously, this dependence on the lapse and shift 
functions is required in order to make the variations of $\UT N$ and 
$N^{a}$ projectable under the Legendre map.  The resulting variations 
under perpendicular diffeomorphisms ($PD$), with descriptor $\xi^{0}$ 
(with $\UT\xi^{0}=t^{-1}\xi^{0}$, which will be useful later), are
\begin{mathletters}\label{trd_trd}
\begin{eqnarray}
    \delta_{PD}[\xi^{0}]\UT N 
    &=&  \UT{\dot\xi}{}^{0} + \UT \xi^{0} N^{a}_{,a}
        - N^{a} \UT \xi^{0}_{,a} \ ,
            \label{dPDN} \\
\delta_{PD}[\xi^{0}] N^{a}
    &=& - Ne^{ab}\xi^{0}_{,b} + N_{,b}e^{ab}\xi^{0} \ .
            \label{dPDNa}
\end{eqnarray}
\end{mathletters}%

The resulting variation of $\OT T^{a}_{i}$ is 
\cite{pons/salisbury/shepley/99b}
\begin{eqnarray}
    \delta_{PD}[\xi^{0}] \OT T^{a}_{i}
    &=& -\xi^{0} \epsilon^{ijk} \Omega^{k}_{\mu}n^{\mu} \OT T^{a}_{j}
        - \UT \xi^{0} \OT T^{b}_{i} \OT T^{a}_{j} K^{j}_{b}
            \nonumber \\
    &&\qquad   + \UT \xi^{0}\OT T^{a}_{i} \OT T^{b}_{j} K^{j}_{b} \ .
            \label{trdvar} 
\end{eqnarray}
We can rewrite the variation of $\OT T^{a}_{i}$ in terms of the 
canonical variables, using the equation of motion (\ref{obstacle}) so 
that
\[  \Omega^{i}_{\mu} n^{\mu} 
    = A^{i}_{\mu} n^{\mu}
        -i N^{-1} T^{ai}N_{,a} \ . 
\]
Also, using equation of motion (\ref{ast}), we find
\begin{eqnarray}
    &&- \UT \xi^{0} \OT T^{b}_{i} \OT T^{a}_{j} K^{j}_{b}
        + \UT \xi^{0}\OT T^{a}_{i} \OT T^{b}_{j} K^{j}_{b} 
            \nonumber \\
    && \qquad =
        -i {\cal D}_{b} (\epsilon^{ijk} 
            \OT T^{b}_{j} \OT T^{a}_{k}\UT \xi^{0} ) 
        + i \epsilon^{ijk}\OT T^{b}_{j} \OT T^{a}_{k}
                {\cal D}_{b} \UT \xi^{0} \ .
\end{eqnarray}
The result is that
\begin{eqnarray}
    \delta_{PD}[\xi^{0}] \OT T^{a}_{i}
    &&=  \xi^{0} \epsilon^{ijk} A^{j}_{\mu}n^{\mu} \OT T^{a}_{k} 
        -i\epsilon^{ijk}N^{-1} \OT T^{bj}
             \OT T^{a}_{k} \xi^{0} {\cal D}_{b}\UT N
                \nonumber \\
    &+& i {\cal D}_{a} (\epsilon^{ijk} \OT T^{b}_{j} \OT T^{a}_{k}
                               \UT \xi^{0} ) 
        + i \epsilon^{ijk}\OT T^{b}_{j} \OT T^{a}_{k} 
               {\cal D}_{b} \UT \xi^{0} \ .
                    \label{dPDTvar}
\end{eqnarray}

The variation of the Ashtekar connection requires a little more work.  
Since under perpendicular diffeomorphisms we will be concerned only 
with on-shell variations (that is, variations of solutions), our task 
is to find the appropriate variations of the four-dimensional Ricci 
rotation coefficients.  We begin with the three-dimensional 
coefficients $\omega^{i}_{a}$, which are constructed from the triad 
and whose variation therefore requires only (\ref{trdvar}).  We showed 
in \cite{pons/salisbury/shepley/99b} that generally
\begin{eqnarray*}
    \delta \omega^{ij}_a
    &=& \UTT g_{a c} \OT T^{b[i} D_{b}\delta \OT T ^{j] c}
        + \OT T^{b [i} \UT t^{j]}_c \UT t^k_a D_{b}
            \delta \OT T^c_{k}
                 \\
    &&+ \UT t^{[i}_b D_{a}\delta \OT T^{j] b}
        + \UT t^k_c  \UT t^{[i}_a \OT T^{j]b} D_{b}
            \delta \OT T^c_{k} \ .
\end{eqnarray*}
Using (\ref{trdvar}) we find
\begin{eqnarray}
    \delta_{PD}[\xi^{0}] \omega^{i}_{a}
    &=& 2 \epsilon^{ijk} T^{b}_{j} D_{[a}K^{k}_{b]} \xi^{0} 
        + D_{a} (\xi^{0} n^{\mu} \Omega^{i}_{\mu}) 
            \nonumber \\
    && - \epsilon^{ijk} T^{b}_{j} \xi^{0}_{,b} K^{k}_{a} \ .
             \label{dPDo}
\end{eqnarray}
Note that (\ref{ddOHM}) demonstrates that
\[  \delta_{S'}[\UT \xi^{0}]  
    = \delta_{PD}[t\UT \xi^{0}] 
    + \delta_{R}[t\UT \xi^{0}n^{\mu}\Omega_{\mu}] \ .
\]

We will calculate the variation of $\Omega^{0i}_{a}$ in 
(\ref{omeg-a0i}) using the expression
\begin{equation}
    K^{i}_{a} := T^{bi}K_{ab} = N T^{bi}\,{}^{4}\Gamma^{0}_{ab} \ .
\end{equation}
The general variation of the four-dimensional Christoffel symbols 
${}^{4}\Gamma^{0}_{ab}$ under a diffeomorphism with descriptor 
$\epsilon^{\mu}$ is
\begin{equation}
    \delta{}^{4}\Gamma^{0}_{bc}
    = - {}^{4}\Gamma^{\sigma}_{bc}\epsilon^{0}_{,\sigma}
        + {}^{4}\Gamma^{0}_{\sigma c}\epsilon^{\sigma}_{,b}
        + {}^{4}\Gamma^{0}_{b \sigma}\epsilon^{\sigma}_{,c}
        + \epsilon^{0}_{,bc}
        + {}^{4}\Gamma^{0}_{bc,\sigma}\epsilon^{\sigma} \ .
\end{equation}
Using methods employed in \cite{pons/salisbury/shepley/99a}, we find
\begin{eqnarray}
    \delta_{PD}[\xi^{0}] K^{i}_{a} 
    &=& -T^{b}_{j}( {}^{3}\!R^{ij}_{ab} 
        + K^{i}_{a} K^{j}_{b} -  K^{i}_{b} K^{j}_{a}) \xi^{0}
            \nonumber \\
    && + (T^{bi} \xi^{0}_{,b} )_{,a} 
        -\epsilon^{ijk}  T^{b}_{j} 
            \xi^{0}_{,b} \omega^{k}_{a} 
                \nonumber \\
    && + \epsilon^{ijk} \xi^{0} n^{\mu} 
            \Omega^{j}_{\mu} K^{k}_{a} \ .
                \label{dPDK}
\end{eqnarray}
Finally, substituting (\ref{dPDo}) and (\ref{dPDK}) into 
\[  \delta_{PD}[\xi^{0}] A^{i}_{a} 
    = \delta_{PD}[\xi^{0}] \omega^{i}_{a}
        + i \delta_{PD}[\xi^{0}] K^{i}_{a} \ ,
\] 
we find that on-shell
\begin{eqnarray}
    \delta_{PD}[\xi^{0}] A^{i}_{a} 
      &=& -i T^{b}_{j} F^{ij}_{ab} \xi^{0}
            \nonumber \\
    &&  -\delta_{R}[\xi^{0}n^{\mu} A_{\mu}
        -i \xi^{0}N^{-1} T^{b}N_{,b}]A^{i}_{a}
            \nonumber \\
    && +\delta_{R}[-iT^{b} \xi^{0}_{,b}]A^{i}_{a} \ .
\end{eqnarray}

We turn finally to the variation of $A^{i}_{0}$.  Results obtained in 
\cite{pons/salisbury/shepley/99b} are
\begin{eqnarray}
    \delta_{PD}[\xi^{0}] \Omega_{0}^{ij}
    &=& -4 \xi^{0} N^{a} D_{[a} K_{b]}^{[i} T^{j]b}
            \nonumber \\
    &&+ 2 N^{b} \xi^{0}_{,a} K^{[i}_{b} T^{j]a}
        +2 N_{,b} \xi^{0}_{,a} T^{b[i}T^{j]a}
            \nonumber \\
    &&+(\Omega_{\mu}^{ij} n^{\mu} \xi^{0})_{,0}
        +2 \xi^{0} n^{\mu} \Omega_{\mu}^{[i} \Omega_{0}^{j]} \ ,
            \label{dPDOmi}
\end{eqnarray}
and
\begin{eqnarray}
    \delta_{D}[\vec \xi] \Omega_{0}^{i} 
    &=& -\epsilon^{ijk}\xi^{a} (K_{a}^{j}T^{bk} N_{,b} 
                                + 2 N T^{bj} D_{[a} K_{b]}^{k}) 
            \nonumber \\
    && - {}^{3}\!R_{ba}^{i} N^{b} \xi^{a}
        + (\xi^{a}\omega_{a}^{i})_{,0} 
            \nonumber \\
    &&  +\epsilon^{ijk}\xi^{a}\omega_{a}^{j} \Omega_{0}^{k}\ .
                \label{deltaOmegac}
\end{eqnarray}
The most efficient calculation of the on-shell variation of 
$\Omega^{0i}_{0}$ is accomplished by proceeding from the expression 
(\ref{omeg-00i}), using the variations (\ref{trd_trd}) and 
(\ref{dPDK}).  For this purpose we also require the variation
\begin{equation}
    \delta_{PD}[\xi^{0}] N_{,a} 
    = -\xi^{0}_{,b} N^{b}_{,a} 
        -\xi^{0}_{,ab} N^{b} 
        - \xi^{0}_{,0} N^{-1} N_{,a} +\xi^{0}_{,0a} \ .
\end{equation}
The result is
\begin{eqnarray}
    && \delta_{PD}[\xi^{0}] \Omega^{0i}_{0} 
            \nonumber \\
    && \qquad = -N^{a} T^{b}_{j} \xi^{0} ( {}^{3}\!R^{ij}_{ab} 
        + K^{i}_{a} K^{j}_{b} -  K^{i}_{b} K^{j}_{a}) 
            \nonumber \\
    && \qquad  \quad + \xi^{0}_{,a} (2 D_{b} N^{[b} T^{a]i} 
        -N T^{a}_{i} T^{b}_{j}K^{j}_{b}
        +N T^{b}_{i} T^{a}_{j}K^{j}_{b}) 
            \nonumber \\
    && \qquad  \quad + \OT T^{bi} (D_{b}\UT\xi^{0})_{,0} 
        +\epsilon^{ijk} \xi^{0} n^{\mu} 
        \Omega^{j}_{\mu} \Omega^{0k}_{0} \ .
            \label{dPDOm0i}
\end{eqnarray}
Using (\ref{dPDOmi}) and (\ref{dPDOm0i}), we deduce that on-shell
\begin{eqnarray}
    && \delta_{PD}[\xi^{0}] A^{i}_{0} 
                \nonumber \\
    && \qquad = -i N^{a} T^{b}_{j} F^{ij}_{ab} \xi^{0} 
                \nonumber \\
    && \qquad\quad + i (T^{bi} \xi^{0}_{,b})_{,0}
        +i \epsilon^{ijk} T^{bj} \xi^{0}_{,b} A^{k}_{0}
                \nonumber \\
    && \qquad\quad + (\xi^{0} n^{\mu} A^{i}_{\mu}
        -i\xi^{0}N^{-1} T^{bi}N_{,b})_{,0} 
                \nonumber \\
    && \qquad\quad + \epsilon^{ijk}(\xi^{0}n^{\mu} A^{j}_{\mu}
            -i\xi^{0}N^{-1} T^{bj}N_{,b}) A^{k}_{0} \label{dPDA1} \\
    && \qquad = -i N^{a} T^{b}_{j} F^{ij}_{ab} \xi^{0} 
                \nonumber \\
     && \qquad\quad - \delta_{R}[\xi^{0}n^{\mu} A_{\mu}
        - i\xi^{0}N^{-1} T^{b}N_{,b}] A^{i}_{0}
               \nonumber \\
    && \qquad\quad + \delta_{R}[-i T^{b} \xi^{0}_{,b}] A^{i}_{0} \ .
           \label{dPDA2}     
\end{eqnarray}

Notice that this variation is not projectable under the Legendre map
due to the presence of time derivatives of the gauge functions
$A^{i}_{0}$, $N$, and $N^{a}$ in the next to last line of
(\ref{dPDA1}).  But fortunately, the final two lines of (\ref{dPDA1})
are a variation under a gauge rotation with descriptor 
\[ \theta^{i} 
    = -\xi^{0}n^{\mu} A^{i}_{\mu}
	+i\xi^{0}N^{-1} T^{bi}N_{,b} \ .
\]
That means we must accompany perpendicular diffeomorphisms with a 
gauge rotation with the descriptor $-\theta^{i}$ to obtain a gauge 
variation which is projectable under the Legendre map.  It is 
significant that on-shell, according to (\ref{obstacle}), 
$-\theta^{i}=\xi^{0}n^{\mu}\Omega^{i}_{\mu}$, so that in the real 
triad sector the required gauge rotation is real, and in fact we 
recover the same projectability condition as in the real triad 
formulation of general relativity \cite{pons/salisbury/shepley/99b}.

Finally we write down the variation of $A^{i}_{0}$ under a spatial 
diffeomorphism.  Since $\Omega^{i}_{\mu}$ and $\Omega^{0i}_{\mu}$ each 
transform as a four-vector under these transformations, the result is 
the usual Lie derivative,
\begin{equation}
    \delta_{D}[\vec \xi] A^{i}_{0} 
    = \dot \xi^{a} A^{i}_{a} +\xi^{a} A^{i}_{0,a} \ .
\end{equation}


\section{Symmetry Generators}
\label{sec:gen}

We now turn to the gauge group itself and the structure and algebra of 
the generators of this group.


\subsection{Group algebra}
\label{subsec:alg}

First, we will find the transformations of non-gauge variables 
generated by each of the secondary constraints.  For this purpose let 
us define
\begin{mathletters}\label{RVSRVS}
\begin{eqnarray}
    R[\xi] 
    &:=& \int d^{3}x\,\xi^i\OT{\cal H}_{i} \ , 
        \label{RVS.R} \\
    V[\vec \xi] 
    &:=& \int d^{3}x\,\xi^a\OT{\cal H}_{a} \ , 
        \label{RVS.V} \\
    S[\UT\xi^{0}] 
    &:=& \int d^{3}x\,\UT\xi^0 \OTT{\cal H}_{0} \ .
        \label{RVS.S}
\end{eqnarray}
\end{mathletters}%
These generators are written at a given time (that is not explicitly
given in the notation).  All brackets associated with them are
equal-time brackets.  These generate gauge rotations, spatial
diffeomorphisms plus associated gauge rotations, and perpendicular
diffeomorphisms plus associated gauge rotations, respectively.  We
have, for example,
\begin{mathletters}
\begin{eqnarray}
    &&\{\OT T^a_i,R[\xi]\}
        =  -\epsilon_{ijk} \xi^j \OT T^a_k 
        := \delta_{R}[\xi] \OT T^a_i \ , \\
    &&\{\OT T^a_i,V[\vec \xi]\}
        = \xi^b_{,b} \OT T^a_i +\xi^b \OT T^a_{i,b}
            - \xi^a_{,b} \OT T^b_i 
            - \xi^b \epsilon_{ijk} A^j_b \OT T^a_k  
                \nonumber \\
    &&\qquad\quad = {\cal L}_{\vec \xi} \OT T^a_i 
            + \delta_{R}[\xi^b A_b] \OT T^a_i \ , 
                \label{dVT} \\
    &&\{\OT T^a_i,S[\UT \xi^{0}]\}
         = -i {\cal D}_{b} (\epsilon^{ijk} 
                   \OT T^{b}_{j} \OT T^{a}_{k}
                                 \UT \xi^{0}) 
                    \nonumber \\ 
    &&\qquad\quad =\delta_{PD}[t \UT \xi^{0}] \OT T^a_i 
                    \nonumber \\
    &&\qquad\quad\quad + \delta_{R}[\xi^{0}  A_{\mu}n^{\mu} 
                -iN^{-1} T^{b}N_{,b}\xi^{0}] \OT T^a_i 
                   \nonumber \\
    &&\qquad\quad\quad -\delta_{R}[ -i \OT T^{b} 
                  {\cal D}_{b} \UT \xi^{0}] \OT T^a_i \ .
                    \label{dST}
\end{eqnarray}
\end{mathletters}%
Thus, according to our discussion following (\ref{dPDA2}), $S[\UT 
\xi^{0}]$ does indeed generate a projected variation.  Notice also 
that we obtain a real projected variation of a real triad if we undo 
the imaginary rotation of the triad due to the imaginary descriptor 
$i\OT T^{b}_{j}{\cal D}_{b}\UT\xi^{0}$ in (\ref{dST}).  The 
generator on non-gauge variables is
\begin{equation}
    S'[\UT\xi^{0}]
    := \int d^{3}x\bigr(\UT\xi^{0}\OTT{\cal H}_{0}
      -i ({\cal D}_{a}\UT\xi^{0})\OT T^{ai}\OT{\cal H}_{i}\bigr) \ .
          \label{SSprime}
\end{equation}
As we noted in the discussion preceding (\ref{projdelta}), in the real 
triad sector this object generates the same variations as the scalar 
generator $S[\UT\xi^{o}]$ in the real triad theory 
\cite{pons/salisbury/shepley/99b}.

It is convenient from a geometrical perspective to define generators 
of non-gauge variables which effect pure spatial diffeomorphisms.  
Using (\ref{dVT}) we deduce that the required generator is
\begin{equation}
    D[\vec \xi] 
    := \int d^{3}x\,\xi^a ( \OT{\cal H}_{a}
        -A^{i}_{a} \OT{\cal H}_{i}) 
    = \int d^{3}x\,\xi^a \OT{\cal G}_{a} \ .
            \label{Dgen}
\end{equation}
This is the real triad sector term we isolated in (\ref{newhc}).

We are now in position to calculate the entire group algebra from the 
transformation properties in configuration-velocity space, projected 
to phase space.  The projections under the Legendre map of the 
variations of the generators are Poisson brackets of generators.  The 
calculations parallel those in 
\cite{pons/salisbury/shepley/99a,pons/salisbury/shepley/99b}, except 
here it is technically simpler, and conceptually rewarding, also to 
calculate the Poisson bracket $\{S[\UT\xi^{0}],S[\UT\eta^{0}]\}$ in 
this manner.  The nonvanishing Poisson brackets are
\begin{mathletters}
\begin{eqnarray}
    \{R[\xi],R[\eta]\}
        &= & -R[[\xi,\eta]] \ , \\
    \{R[\xi],D[{\vec \eta}]\}
        &=& -R[{\cal L}_{\vec\eta}\xi] \ , \\
    \{D[{\vec \xi}],D[{\vec \eta}]\}
        &=& -D[{\cal L}_{\vec\eta}{\vec \xi}] 
            = D[[{\vec \xi},{\vec\eta}]] \ , \\
    \{S[{\UT \xi^{0}}],D[{\vec\eta}]\}
        &=& - S[{\cal L}_{\vec \eta}\UT\xi^{0}] \ , \\
    \{S[{\UT \xi^{0}}],S[{\UT \eta^{0}}] \}
        &=& V[{\vec\zeta}] \ ,
            \label{veczeta}
\end{eqnarray}
\end{mathletters}%
where in (\ref{veczeta})
\begin{equation}
    \zeta^a 
        := (\UT\xi\partial_{b}\UT\eta 
            - \UT\eta\partial_{b}\UT\xi) \OTT e^{ab} \ .
\end{equation}

It will be useful in constructing the final complete gauge generators 
to have the algebra of the set $R$, $V$, and $S$.  Using the brackets 
above the remaining non-vanishing brackets are
\begin{mathletters}
\begin{eqnarray}
    \{V[{\vec \xi}],V[{\vec \eta}]\}
        &=& V[[{\vec \xi},{\vec \eta}]] 
            -R[\xi^a \eta^b F_{ab}] \ , \\
    \{S[{\UT \xi^{0}}],V[{\vec \eta}]\}  
        &=& -S[{\cal L}_{\vec \eta} \UT \xi^{0}]
            -R[-i \OT T^b_{j} F_{a b}^{i j}\eta^{a} \UT \xi^{0}] \ ,
\end{eqnarray}
\end{mathletters}
where for clarity we use the notation $R[\xi^{i}]$ in the last 
equation instead of $R[\xi]$ as in (\ref{RVS.R}). 


\subsection{Complete symmetry generators}
\label{subsec:symgens}

The canonical Hamiltonian in terms of the generators takes the form
\[  H_{\rm c} = \int d^{3}x\,N^A {\cal H}_A 
    =: N^{A}{\cal H}_{A}\ ,
\]
where we define 
\[  N^A :=\{\UT N, N^a, -A^i_0\}\ ,\
    {\cal H}_A :=\{\OTT{\cal H}_0, \OT{\cal H}_a,\OT{\cal H}_i\}\ ,
\]
and where spatial integrations over corresponding repeated capital 
indices are assumed.  It was shown in \cite{pons/salisbury/shepley/97} 
that the complete symmetry generators then take the form
\begin{equation}
    G(t) = \xi^{A} G^{(0)}_{A} + \dot \xi^{A} G^{(1)}_{A}\ ;
           \label{Gxi}
\end{equation}
the descriptors $\xi^{A}$ are arbitrary functions:
\[  \xi^{A} = \{\UT\xi^{0},\xi^{a},\xi^{i}\} \ .
\]
The simplest choice for the $G^{(1)}_{A}$ are the primary constraints 
$P_{A}$,
\[  P_{A} := \{\OTT P,\OT P_{a},-\OT P_{i}:=-\OT\pi^{0}_{i}\}\ ,
\]
with the result that
\begin{equation}
    G[\xi^{A}] = P_{A} \dot\xi^{A} 
       + ({\cal H}_{A} 
       + P_{C''}N^{B'}{\cal C}^{C''}_{AB'})\xi^{A}\ , 
           \label{GGGGG}
\end{equation}
where the structure functions are
\begin{equation}
    \{ {\cal H}_{A},{\cal H}_{B'} \}
    =: {\cal C}^{C''}_{AB'} {\cal H}_{C''}\ .
\end{equation}

Using the brackets calculated in the previous Section we read off the 
following non-vanishing structure functions:
\begin{eqnarray*}
    C^a_{0'0''}
    &=& \OTT e^{ab}\Big(-\delta^3(x-x')\partial''_b 
            \delta^3 (x-x'')  \\
      &&\qquad +\delta^3 (x-x'')\partial'_b\delta^3 (x-x')\Big)\ , \\
    C^a_{b' c''}
    &=&-\delta^3 (x-x') \partial''_b\delta^3 (x-x'') \delta^a_c  \\
      &&+ \delta^3 (x-x'') \partial'_c \delta^3 (x-x')
          \delta^a_b\ , \\
    C^i_{j' k''}
    &=&-\epsilon^{ijk} \delta^3 (x-x') \delta^3 (x-x'') \ ,\\
    C^0_{0' a''}
    &=&\delta^3 (x-x'') \partial'_a\delta^3 (x-x')  \\
      &&-\delta^3 (x-x') \partial''_a\delta^3 (x-x'') \ , \\
    C^i_{0' a''}
    &=& i \OT T^{b}_{j} F^{ij}_{ab} \delta^3 (x-x')
          \delta^3 (x-x'') \ , \\
    C^i_{a' b''}
    &=& -F^i_{ab} \delta^3 (x-x') \delta^3 (x-x'') \ . \\
\end{eqnarray*}
With the use of the structure functions derived above, we obtain the 
following generators, denoted by $G_{R}[\xi]$, $G_{V}[{\vec\eta}]$, 
and $G_{S}[\UT\zeta^{0}]$.  These generate, respectively, gauge 
rotations, spatial diffeomorphisms, and perpendicular diffeomorphisms 
(plus associated gauge rotations in the last two cases):
\begin{mathletters}
\begin{eqnarray}
    G_{R}[\xi]
    &:=& \int d^{3}x \bigr(\OT{\cal H}_i \xi^{i}
        -\OT P_i (\dot{\xi}^{i}+ \epsilon_{ijk}\xi^j A^k_0)
        \bigr) \ , 
                \label{GRGR}  \\
    G_{V}[{\vec \eta}]
    &:=& \int d^{3}x\, \bigr( \OT{\cal H}_{a}\eta^{a}
          +\OTT P (\UT N_{,a}\eta^a - \UT N \eta^a_{,a})
              \nonumber \\
    &&\qquad    +\OT P_a ( \dot{\eta}^{a} + N^a_{,b} \eta^b
                   - N^b \eta^a_{,b})
                       \nonumber \\
    &&\qquad    + \OT P_i(F^i_{ab}\eta^a N^b 
          +i F^{ij}_{ab} \OT T^{b}_{j} \UT N \eta^{a})\bigr) \ ,
                         \\
    G_{S}[\UT \zeta^{0}]
    &:=& \int d^{3}x\, \bigr(\OTT{\cal H}_{0}\UT{\zeta}^{0}
        + \OTT P(\UT{\dot\zeta}^0- N^a \UT \zeta_{,a}^{0}
            +N^a_{,a} \UT \zeta^{0})
                \nonumber \\
    &&\qquad + \OT P_a(\UT N_{,b} \UT\zeta^{0}\OTT e^{ab}
            - \UT N \UT \zeta_{,b}^{0}\OTT e^{ab}) 
                \nonumber \\
    &&\qquad   -i \OT P_{i} N^{a}\OT T^{b}_{j}
                           F^{ij}_{ab} \UT\zeta^{0}
               \bigr) \ .
                    \label{GSGS}
\end{eqnarray}
We wish to emphasize the following point: Notice that the variation of 
$A^{i}_{0}$ generated by $G_{S}[\UT\xi^{0}]$ is, using (\ref{dPDA2}),
\begin{eqnarray*}
    &&\{A^{i}_{0},\,G_{S}[\UT\xi^{0}]\}
        =-i N^{a} \OT T^{b}_{j} F^{ij}_{ab}\UT\xi^{0} \\
    &&\qquad =\delta_{PD}[\xi^{0}] A^{i}_{0} 
        +\delta_{R}[\xi^{0}n^{\mu} A_{\mu} 
                - i \xi^{0}N^{-1} T^{b}N_{,b}] A^{i}_{0} \\
    &&\qquad\quad - \delta_{R}[-i T^{b} \xi^{0}_{,b}] A^{i}_{0} \ .
\end{eqnarray*}
The second term removes the offending time derivatives of gauge 
variables, so that the first two variations taken together are 
projectable.  The third variation is projectable, and in fact when 
combined with the variation generated by $G_{S}[\UT\xi^{0}]$ produces 
a variation which conserves the reality of real triads, as we noted in 
defining the generator $S'[\UT \xi^{0}]$ in (\ref{SSprime}).  The 
general relation is
\begin{eqnarray}
    && \delta_{PD}[\xi^{0}] 
        +\delta_{R}[\xi^{0}n^{\mu} A_{\mu} 
                - i \xi^{0}N^{-1} T^{b}N_{,b}]
                    \nonumber \\
    && \qquad = \{-,\,G_{S}[\xi^{0}]\}
            +\{-,\,G_{R}[i T^{b} \xi^{0}_{,b}]\}
                    \nonumber \\
    && \qquad =: \{-,\,G_{S'}[\UT\xi^{0}]\} \ .
        \label{GS'}
\end{eqnarray}
Note that the secondary constraint term in $G_{S'}$ is just 
(\ref{SSprime}).
    
Finally, we use the generators above to construct $G_{D}[\vec\xi]$, 
the complete generator of spatial diffeomorphisms with descriptor 
$\vec\xi$.  Refer to (\ref{Dgen}); the generator is evidently, using 
the equation of motion (\ref{eq-a}),
\begin{eqnarray}
    && G_{D}[\vec \xi] = G_{V}[{\vec \xi}] - G_{R}[A_{a}\xi^{a}] 
                  \nonumber \\
    &&\qquad = \int d^{3}x \bigr(\OT{\cal G}_{a}\xi^{a}
        +\OTT P(\UT N_{,a} \xi^a-\UT N \xi^a_{,a})
            \nonumber \\
    &&\qquad\qquad \qquad  +\OT P_a(\dot{\xi}^{a}+N^a_{,b} \xi^b 
                   -N^b \xi^a_{,b})
            \nonumber \\
    &&\qquad \qquad \qquad  +\OT P_i(\dot{\xi}^{a}A^{i}_{a} 
                    +\xi^{a} A^{i}_{0,a})
                \bigr) \ .
                     \label{GDGD}
\end{eqnarray}
\end{mathletters}%


\subsection{The Hamiltonian and rigid time translation}
\label{subsec:global}
 
Now that we have the complete set of generators, we can reconstruct 
the Hamiltonian, recognizing that rigid (in the sense of advancing by 
the same infinitesimal parameter on each constant-time hypersurface) 
translation in time is a diffeomorphism implemented on restricted 
members of equivalence classes of solution trajectories.  We take as 
given explicit spacetime functions $\UT\xi^{0}$ and $\xi^{a}$.  We 
restrict our attention to solutions for which 
$t\UT\xi^{0}=N\delta\tau$ and $\xi^{a}=N^{a}\delta\tau$ for some 
infinitesimal parameter $\delta\tau$.  However, we recall that
\[  \int d^{3}x\,  \UT \xi^{0} \OTT {\cal H}_{0} 
    + \int  d^{3}x\, \xi^{a} \OT {\cal G}_{a}
\] 
does not generate a pure diffeomorphism.  We must subtract the 
additional gauge rotation generated by 
$\int{}d^{3}x\,\UT\xi^{0}\OT{\cal H}_{0}$.  According to (\ref{dST}) 
the descriptor of this gauge rotation is
\begin{equation}
   \xi^{0}  A^{i}_{\mu}n^{\mu}  -\xi^{0}iN^{-1} T^{bi}N_{,b}  
    + i \OT T^{bi} {\cal D}_{b} \UT \xi^{0} \ .
\end{equation}
When we restrict this descriptor to those solutions for which 
$\xi^{0}=N\delta\tau$ and $\xi^{a}=N^{a}\delta\tau$, the descriptor 
becomes $A^{i}_{0}\delta \tau-A^{i}_{a}N^{a}\delta \tau$.  We deduce 
that the required Hamiltonian is
\begin{equation}
    H = \int \, d^{3}x \,\UT N \OTT{\cal H}_{0} 
        + \int \, d^{3}x\,N^{a} \OT{\cal H}_{a} 
        -\int \, d^{3}x \,A^{i}_{0} \OT{\cal H}_{i} \ , 
           \label{ham}
\end{equation}
where we have used the fact that 
$\OT{\cal G}_{a}=\OT{\cal H}_{a}-A^{i}_{a}\OT{\cal H}_{i}$.  The 
Hamiltonian in (\ref{ham}) is coincident with the canonical 
Hamiltonian (\ref{cancan})!
 
The gauge variables $N,N^{a},A^{i}_{0}$ in (\ref{ham}) are now to be 
thought of as arbitrarily chosen but explicit functions of spacetime.  
This object (\ref{ham}) will then generate a time translation, which 
is rigid in the sense of having the same constant value $\delta \tau$ 
on each equal-time hypersurface, but only on those members of 
equivalence classes of solutions for which the dynamical variables 
$N,N^{a},A^{i}_{0}$ have the same explicit functional forms.  On all 
other solutions the corresponding variations correspond to more 
general diffeomorphism and gauge transformations.

In fact, as we pointed out in \cite{pons/salisbury/shepley/99a}, every 
generator $G[\xi^{A}]$ in (\ref{GGGGG}) with $\xi^{0} > 0$ may be 
considered to be a Hamiltonian in the following sense: 
\[  G[\xi^{A}] 
    = G_{R}[\xi] + G_{D}[{\vec\xi}] + G_{S}[\xi^{0}]
\]
generates a global time translation on those solutions which have
\begin{mathletters} \label{HAM.DESCRIP}
\begin{eqnarray}
    N \delta \tau= \xi^{0}\ ,  
            \label{zeta.0} \\
     N^{a} \delta \tau= \xi^{a}\ ,
            \label{eta.a} \\
    (-A^{i}_{0} + A^{i}_{a} N^{a}) \delta \tau = \xi^{i} \ .
    \label{xi.i}
\end{eqnarray}
\end{mathletters}%
We have already demonstrated this fact for the non-gauge variables, 
and it is instructive to verify the claim for the gauge variables $N$, 
$N^{a}$, and $A^{i}_{0}$.  The demonstration for $N$ and $N^{a}$ is 
given in \cite{pons/salisbury/shepley/99a}.  Substituting 
(\ref{HAM.DESCRIP}) into (\ref{GRGR},\ref{GSGS},\ref{GDGD}), we have
\begin{mathletters} \label{HAM.var}
\begin{eqnarray}
   \delta A^{i}_{0} &=& \big(-(-A^{i}_{0} + A^{i}_{a} N^{a})_{,0}
       -\epsilon^{ijk}(-A^{j}_{0} + A^{j}_{a} N^{a}) A^{k}_{0}
            \nonumber \\
        && +A^{i}_{a}\dot N^{a} + N^{a} \dot A^{i}_{a}
            +i F^{ij}_{ab} T^{b}_{j} N^{a} N 
                \nonumber \\
        && +\epsilon^{ijk} N^{a} A^{j}_{a} A^{k}_{0}
         -i F^{ij}_{ab} T^{b}_{j} N^{a} N \big) \delta \tau 
		        \nonumber \\
         &&= \dot A^{i}_{0} \delta \tau\ .
    \label{dA}
\end{eqnarray}
\end{mathletters}%

\subsection{Finite real gauge transformations}
\label{subsec:fingen}

We close this Section by noting that the arguments presented in 
Section \ref{sec:real} demonstrating the preservation of reality 
conditions under time evolution apply almost unaltered to finite 
arbitrary symmetry transformations.  The only restrictions which must 
be placed on the descriptors $\xi^{i}$ and $\xi^{a}$ are that they be 
real.  The triad reality condition implies in addition that we must 
employ the generator $G_{S'}[\UT\xi^{0}]$, defined in (\ref{GS'}), 
instead of $G_{S}[\UT\xi^{0}]$, defined in (\ref{dST}), and the 
descriptor $\UT\xi^{0}$ must be real.  Then we find, as in 
(\ref{stabt}), with the simple substitutions 
$\UT{N}\rightarrow\UT\xi^{0}$, $N^{a}\rightarrow\xi^{a}$, that
\begin{equation} 
    \Im\{\OT T^{a}_{i},\, G_{S'}[\UT \xi^{0}] \} = 0 \ ,
\end{equation}
when $\UT\xi^{0}$ is real.  The next and higher levels of reality 
stabilization are satisfied, just as in Section \ref{sec:real}, with 
the substitutions $\UT{N}\rightarrow\UT\xi^{0}$, 
$N^{a}\rightarrow\xi^{a}$.

The complete infinitesimal gauge generator which respects the triad 
reality condition is
\begin{equation}
     G_{\rm real}[\xi^{A}] 
     := G_{R}[\xi] + G_{D}[\vec\xi] + G_{S'}[\UT\xi^{0}] \ ,
\end{equation}
where $\xi^{A}$ are real (if one has only the metric reality 
conditions, then only $\vec\xi$ and $\UT\xi^{0}$ need be real).  
Finally, the finite real generator (which complies with the triad 
reality conditions), for finite parameter $\tau$, is
\[  {\cal T} \exp\left(\int^{t_{0}+\tau}_{t_{0}} dt \,
                \{ - , G_{\rm real}[\xi^{A}] \}\right)\ .
\]


\section{Counting the degrees of freedom} 
\label{sec:count}


\subsection{With the metric reality conditions}
\label{subsec:count.metric}

Let us again stress the relevant role of the variables in 
(\ref{them}):
\[  M_{ab} := -i t^{i}_{a}(A_b^{i} - \omega_b^{i})\ .
\]
We substitute 
\begin{equation}
    A_a^{i}=\omega^{i}_{a} + iT^{b}_{i}M_{ba}
        \label{a(m)}
\end{equation}
into the constraints (\ref{seconda}) and (\ref{secondb}) (remember 
that the content of (\ref{secondc}) is the condition that $M_{ab}$ be 
symmetric).  We get, for (\ref{seconda}) (${}^{3}\!R$ is the 
three-Ricci scalar),
\begin{equation}
    {}^{3}\!R + (M^{a}_{a})^{2} - M^{a}_{b}M^{b}_{a} = 0\ , 
        \label{hamconst}
\end{equation}
and for (\ref{secondb}),
\begin{equation}
    {}^{3}\nabla_{a}M^{b}_{b} - {}^{3}\nabla_{b}M^{b}_{a} = 0\ .
         \label{momconst}
\end{equation}
These are the standard scalar and vector constraints for canonical ADM 
general relativity \cite{arnowitt/deser/misner/62}.  This is an 
expected result, because $M_{ab}$ gives, according to (\ref{k=g}), the 
initial values for the components of the extrinsic curvature.

The initial data are, therefore: $N$, $N^{a}$, $M_{ab}$, all real with 
$M_{ab}$ symmetric, and $t^{i}_{a}$, $A^{i}_{0}$, complex.  Thus we 
are implementing the constraints (\ref{secondc}) and the secondary 
reality condition (\ref{2ndreal}).  $A^{i}_{a}$ is then determined by 
(\ref{a(m)}).  This amounts to  $1+3+6+2\times(9+3) = 34$ real pieces 
of data.  But $t^{i}_{a}$ must satisfy the 6 restrictions coming from 
the first metric reality condition (\ref{pmrc}), and both $M_{ab}$ and 
$t^{i}_{a}$ must fulfill the 4 constraints (\ref{hamconst}) and 
(\ref{momconst}).  The number of independent real pieces of data is 
then $34-6-4=24$.

Now let us turn to the gauge freedom.  We have the 4 generators 
corresponding to the space-time diffeomorphisms and the 6 generators 
for $SO(3,C)$, three for real rotations and three for imaginary 
rotations.  This totals 10 generators.  All these generators, as we 
have seen in the previous Section, contain primary and secondary first 
class constraints.  This means that we must spend 2 gauge fixing 
constraints for each generator---see, for example, 
\cite{pons/shepley/95} for the theory of gauge fixing.  Hence we must 
produce $2\times10=20$ gauge fixing constraints to eliminate fully the 
unphysical degrees of freedom.  The final counting of physical degrees 
of freedom is therefore $24-20=4$.  This is the standard number of 
degrees of freedom of general relativity.


\subsection{With the triad reality conditions}
\label{subsec:count.triad}

Now the initial data are: $N$, $N^{a}$, $M_{ab}$, $t^{i}_{a}$, and 
${\Re}A^{i}_{0}$, all real with $M_{ab}$ symmetric.  In this way we 
have already implemented the primary and secondary triad reality 
conditions.  $A^{i}_{a}$ is determined, as before, by (\ref{a(m)}), 
and the imaginary part of $A^{i}_{0}$ is determined by 
(\ref{sectrc2}).  This amounts to $1+3+6+9+3=22$ real pieces of data.  
But $t^{i}_{a}$ and $M_{ab}$ are still constrained to satisfy the 4 
ADM constraints (\ref{hamconst}) and (\ref{momconst}).  The number of 
independent real data is then $22-4=18$.

Now let us turn to gauge freedom.  We have the 4 generators 
corresponding to the spacetime diffeomorphisms and the 3 generators 
that are left after reducing $SO(3,C)$ to $SO(3,R)$ in order to 
preserve (\ref{sectrc2}).  This totals 7 generators.  As we have 
mentioned above, we must introduce 2 gauge fixing constraints for each 
generator.  The final counting of physical degrees of freedom is, 
again, $18-14=4$.


\section{conclusions}
\label{sec:conc}

In this paper we have given a full account of two issues concerning 
the complex Ashtekar approach to canonical gravity: the nature of the 
gauge group and the implementation of reality conditions.  We have 
solved the problem of the projectability of the spacetime 
diffeomorphism transformations from configuration-velocity space to 
phase space; we have constructed the complete set of canonical 
generators of the gauge group in phase space (which includes the gauge 
variables); and we have verified that they indeed generate the 
projected gauge transformations obtained from configuration-velocity 
space.  This result proves that the canonical formalism is capable of 
displaying all the gauge structure of the theory, including the time 
diffeomorphisms, and in particular it proves that the gauge group in 
configuration-velocity space is the same as in phase space---the only 
difference is a matter of a convenient basis for the generators.

The gauge rotations which must be added to spacetime diffeomorphisms 
in achieve projectability differ somewhat from the Einstein-Yang-Mills 
case (see \cite{pons/salisbury/shepley/99a}).  The difference is due 
to the fact that the Ashtekar connection is not a manifest spacetime 
one-form under diffeomorphisms for which the descriptor 
$\epsilon^{0}_{,a}\ne0$.

The full projectable transformation group must be interpreted as a 
transformation group on the space of solutions of the equations of 
motion.  The pullback of variations of $A^{i}_{a}$ from phase space to 
configuration-velocity space yields variations 
$\delta\dot{\OT{T}}{}^{a}_{i}$ which only coincide on-shell with 
${d\over{}dt}\OT{T}{}^{a}_{i}$.  However, if we use only the pullback 
of the variations of the configuration variables, ignoring the 
pullback of momentum variables, the resulting variation of the 
Lagrangian is a divergence (note that we have been ignoring boundary 
terms---for a discussion of the algebra of spatial diffeomorphisms 
including boundary terms, see \cite{soloviev/97}).  These pullbacks 
yield Noether Lagrangian symmetries.  For details see 
\cite{pons/salisbury/shepley/99a,gracia/pons/99}

This restriction to solution trajectories is intimately related to our 
demonstration that all $G[\xi^{A}]$ generators (with $\UT\xi^{0}>0$) 
can be interpreted as Hamiltonians (for time evolution, in the sense 
discussed in Section \ref{sec:gen}).

Since the complex character of the Ashtekar connection introduces the 
issue of reality conditions, we have first produced a general 
theoretical framework for the stabilization algorithm for these 
conditions.  We showed that there are striking differences from 
Dirac's method of stabilization of constraints (reality conditions are 
not constraints in the Dirac sense).  For instance, the calculation 
that shows that the stabilization procedure has been completed is 
typically not nearly as straightforward as in the Dirac case.

Our display of the reality conditions for Ashtekar's formulation is 
not new, but we present a rigorous proof, based on the stabilization 
algorithm, that the set of reality conditions and the algorithmic 
computation are complete.  Also, in the case of the triad reality 
conditions, we showed that the stabilization algorithm implies the 
partial determination of some of the arbitrary functions (actually, 
the determination of their imaginary parts) in the Dirac Hamiltonian 
$H_D$.  We have proved that the the reality conditions are consistent 
with the gauge group.

We note two links between the triad reality conditions and the 
canonical generators associated with projectable diffeomorphisms.  
First, the form of our generator (\ref{Dgen}) for spatial 
diffeomorphisms of the nongauge variables is the same as the form of 
the generator (\ref{ha}) dictated by the triad reality conditions.  In 
contrast, in the Einstein-Yang-Mills case 
\cite{pons/salisbury/shepley/99a}, the form of this generator was more 
a matter of convenience than necessity.  Second, the form of the 
canonical Hamiltonian in (\ref{newhc}) was suggested by triad reality 
conditions.  When $\UT N$ is replaced by $\UT\xi^{0}$ in the third 
term in the integrand, one obtains the generator (\ref{SSprime}) of 
the canonical version of the perpendicular diffeomorphisms---when a 
rotation is subtracted to make these diffeomorphisms projectable; this 
rotation cancels the next to last term in (\ref{dPDA2}).  In fact, the 
rotation which is subtracted is identified as being a real rotation 
within the triad reality conditions (see \ref{sectrc2}).

Finally, we presented the counting of degrees of freedom, either under 
the metric reality conditions or the triad reality conditions.  We 
showed this number matches the standard number of degrees of freedom 
of general relativity.

We feel that this work provides a new understanding of spacetime 
diffeomorphisms in the full (that is, including the gauge variables) 
complex canonical formalism of Ashtekar for gravity.  We expect that 
implications for an eventual quantum theory of gravity will include 
insights into the problem of time in such a theory.  We will be 
investigating these ideas further.


\section*{Acknowledgments}

JMP and DCS would like to thank the Center for Relativity of The 
University of Texas at Austin for its hospitality.  JMP acknowledges 
support by CICYT, AEN98-0431, and CIRIT, GC 1998SGR, and wishes to 
thank the Comissionat per a Universitats i Recerca de la Generalitat 
de Catalunya for a grant.  DCS acknowledges support by National 
Science Foundation Grant PHY94-13063.  We also wish to thank the 
referee for carefully reading this paper and suggesting improvements.




\end{multicols}

\end{document}